\documentclass[final,5p,times,twocolumn]{elsarticle}

\usepackage{graphicx}
\usepackage{epstopdf}
\usepackage{epsfig}
\usepackage{amssymb,amsmath,stmaryrd,tabularx}
\usepackage{wrapfig}
\usepackage{bm}
\usepackage[T1]{fontenc}
\usepackage[center]{subfigure}
\usepackage[toc,page]{appendix}
\usepackage{float}
\usepackage{booktabs}
\usepackage{array}
\usepackage{blindtext}
\usepackage{amsthm}
\usepackage[normalem]{ulem}
\usepackage[percent]{overpic}
\usepackage{multirow}
\usepackage[colorlinks=true, allcolors=blue]{hyperref}
\usepackage{amsfonts}
\usepackage{bbm}
\usepackage{ulem}
\usepackage[percent]{overpic}
\usepackage{textgreek}
\usepackage{changes}
\usepackage{widetext}

\newcommand{\kv}{\mathbf{k}}

\newcommand{\bfdelta}{\textbf{\textdelta}}
\newcommand{\epsc}{\eps_c}
\newcommand{\eps}{\varepsilon}
\newcommand{\bigO}{\mathcal{O}}
\newcommand{\ie}{i.e., }
\newcommand{\rmd}{\,\mathrm{d}}
\newcommand{\rmT}{\mathrm{T}}
\newcommand{\g}{|\pmb{g}|}
\renewcommand{\i}{\mathbbm{i}}
\newcommand{\calF}{\mathcal{F}}
\newcommand{\calG}{\mathcal{G}}
\newcommand{\calH}{\mathcal{H}}
\newcommand{\calK}{\mathcal{K}}
\newcommand{\calL}{\mathcal{L}}
\newcommand{\calLs}{\calL_{_{\scriptstyle\calS}}}
\newcommand{\calM}{\mathcal{M}}
\newcommand{\calS}{\mathcal{S}}
\newcommand{\dS}{\delta_\calS}
\newcommand{\nablaf}{\nabla}
\newcommand{\nablas}{\nabla_{\!\calS}}
\newcommand{\lap}{\nabla^2}
\newcommand{\lapbel}{\nabla_{\!\calS}^2}
\newcommand{\dt}{\partial_t}
\newcommand{\dx}{\partial_x}
\newcommand{\dy}{\partial_y}
\newcommand{\no}{{\overline{\psi}}}

\newcommand{\sPFCnospace}{sPFC}
\newcommand{\sAPFCnospace}{sAPFC}
\newcommand{\sPFC}{\sPFCnospace$\ $}
\newcommand{\sAPFC}{\sAPFCnospace$\ $}

\newcommand{\scal}[2]{\left\langle #1, #2 \right\rangle}

\setcitestyle{authoryear,round}

\begin{document}

\title{Mesoscale modeling of deformations and defects in thin crystalline sheets}

\author[TUD]{Lucas Benoit-{}-Mar\'echal}
\author[TUD]{Ingo Nitschke}
\author[TUD,DCMS]{Axel Voigt}
\author[TUD,DCMS]{Marco Salvalaglio\corref{corrauth}}
\cortext[corrauth]{Corresponding author}
\ead{marco.salvalaglio@tu-dresden.de}
\address[TUD]{Institute of Scientific Computing, TU Dresden, 01062 Dresden, Germany}
\address[DCMS]{Dresden Center for Computational Materials Science (DCMS), TU Dresden, 01062 Dresden, Germany}

\begin{abstract}
We present a mesoscale description of deformations and defects in thin, flexible sheets with crystalline order, tackling the interplay between in-plane elasticity, out-of-plane deformation, as well as dislocation nucleation and motion. Our approach is based on the Phase-Field Crystal (PFC) model, which describes the microscopic atomic density in crystals at diffusive timescales, naturally encoding elasticity and plasticity effects. In its amplitude expansion (APFC), a coarse-grained description of the mechanical properties of crystals is achieved. We introduce surface PFC and surface APFC models in a convenient height-function formulation encoding deformation in the normal direction. This framework is proven consistent with classical aspects of strain-induced buckling, defect nucleation on deformed surfaces, and out-of-plane relaxation near dislocations. In particular, we benchmark and discuss the results of numerical simulations by looking at the continuum limit for buckling under uniaxial compression and at evidence from microscopic models for deformation at defects and defect arrangements, demonstrating the scale-bridging capabilities of the proposed framework. Results concerning the interplay between lattice distortion at dislocations and out-of-plane deformation are also illustrated by looking at the annihilation of dislocation dipoles and systems hosting many dislocations. With the novel formulation proposed here, and its assessment with established approaches, we envision applications to multiscale investigations of crystalline order on deformable surfaces.
\end{abstract}

\maketitle

\section{Introduction}
Thin crystalline sheets are ubiquitous in both natural and manufactured systems. Because of their dimensionality, they display fascinating and often unexpected mechanical properties.
We here focus on atomically thin crystalline 2D materials, e.g.~\cite{Hu2012review,Lehtinen2013,Felton2014,SydneyGladman2016}. A prominent example is graphene, which has been widely studied in the last decade for the peculiar interplay between morphology and mechanical (and electronic) properties it displays, e.g.,~\cite{Lehtinen2013,Pereira2010,Liu2011,Warner2012}. Nonetheless, the general mechanical properties resulting from crystalline order on deformable surfaces also hold for other crystalline structures and colloids, e.g.,~\cite{androulidakis2018tailoring,khanra2022controlling}. Due to the possibility of out-of-plane deformations, these systems exhibit unique behaviors in terms of elastic response to stresses, defect nucleation, defect-induced deformations, and plastic relaxation in general. We will focus on these general properties.

The study of mechanical properties in materials largely benefits from mesoscale approaches, in particular concerning plasticity and defect-induced deformations, e.g.~\cite{Elder2002,bulatov2006computer,kubin2013dislocations,rollett2015understanding}. Bridging microscopic and continuum length scales is necessary to simultaneously account for long-range elastic fields, localized defect-induced effects, and defect-defect interactions. 
In the context of deformable crystalline sheets, effects related to out-of-plane displacement and local curvature add up to this complexity \citep{Lehtinen2013}. Successful modeling approaches dealing with an atomistic description of crystalline sheets are, e.g.,~\cite{Jeong2008,Ariza2010,Cui2020,Torkaman-Asadi2022}. On the other side, there are also continuum descriptions, e.g.,~\cite{Seung1988,Witten2007,Guinea2008,Zhang2014a,Roy2018,singh2022interaction,Singh2022}. Attempts to couple these two descriptions in a multiscale fashion have also been proposed, e.g.,~\cite{Zhang2014a}. 
In this work, we present instead a mesoscale framework that, within one single model formulation, allows for the description of lattice deformations, defects, and comprehensive morphological evolution of crystalline sheets. To that end, we leverage the capability of the so-called phase field crystal (PFC) models~\citep{Elder2002} in a coarse-grained formulation, retaining microscopic details while inspecting large length-scale behaviors.

The PFC model describes lattice structures over relatively large (diffusive) timescales \citep{Elder2002,Elder2004,Emmerich2012} by defining an energy functional minimized by a periodic field $\psi$---approximating the atomic density in crystalline arrangements---and using a conservative gradient flow for its evolution. Despite being a minimal, phenomenological model, such an approach can be derived from classical and dynamical density functional theory \cite{Elder2007,Teeffelen2009,Emmerich2012,Archer2019,teVrugt2020,teVrugt2023}. It has been successfully applied to study several phenomena of crystalline materials self-consistently, including elasticity and plasticity \citep{Emmerich2012,Berry2014,Skaugen2018,Skogvoll2022pfc}. The free energy can be written as Hooke's law upon small deformations of $\psi$ \citep{ElderPRE2010,Heinonen2014}, while defects such as dislocations and grain boundaries naturally emerge as a result of energy minimization for mismatched/tilted crystals \citep{Elder2002}.
In particular, PFC models have been used for the systems targeted in this work, namely crystalline arrangements on curved surfaces \citep{Backofen2010,Backofen2011,Aland2012,Kohler2016,Elder2021}. However, the spatial resolution required by the PFC model limits numerical investigations of large systems, and stress/strain fields are accessible only upon numerical coarse-graining. 

By focusing on the complex amplitudes of $\psi$, a spatially coarse-grained version of the PFC model, the amplitude expansion (APFC) model, has been introduced \citep{Goldenfeld2005, Athreya2006,Salvalaglio2022ov}. This model describes crystalline arrangements by fields (the complex amplitudes) varying over larger length scales while retaining relevant microscopic information on the lattice symmetry. Such an approach has been shown to describe lattice deformations well \citep{Heinonen2014,Huter2016,Heinonen2016,Salvalaglio2019,Salvalaglio2020plastic} and enables self-consistent investigations of crystalline systems with dislocations at relatively large length scales \citep{Salvalaglio2021,Jreidini2021}. Compared to its microscopic counterpart, the APFC model also gives direct access to the stress field as a function of the amplitudes \citep{Salvalaglio2019}. In short, this approach can be seen as a classical phase field model describing phases and interfaces that includes elasticity and dislocations \citep{Salvalaglio2022ov}. However, formulations on curved surfaces have not been proposed so far.

In this work, we fill this gap and introduce a mesoscale framework based on the APFC model to describe deformations of thin crystalline sheets at relatively large scales---approaching macroscopic descriptions---while retaining information on the microscopic length scales and defects. We propose a surface APFC (\sAPFCnospace) model based on a convenient height formulation, whereby the equations are solved in the flat space by encoding the height of the surface with respect to a nominal reference as a variable of the governing equations. To this end, we first devise a novel surface PFC (\sPFCnospace) model that extends the current state of the art \citep{Elder2021} by retaining higher-order terms and considering deformations along the surface normal, thus dispensing with the small height-gradient limit required when only vertical deformations are assumed. 
Its coarse-grained counterpart then naturally ensues. The resulting approach is compatible with the basic assumption entering the derivation of the APFC model---namely, the decomposition of the atomic density in terms of Fourier modes---and allows for a convenient quasi-2D description of surfaces in 3D space. With this model, we study representative cases of deformations of crystalline sheets, both purely elastic and in the presence of dislocations. We assess our description by comparisons with microscopic approaches in terms of dislocation nucleation on deformed surfaces. The scale-bridging nature of the APFC model is also shown by examining the link to macroscopic models, such as continuum elasticity and F\"oppl-von K\'arm\'an (FvK) equations for elastic regimes.

The manuscript is organized as follows. In Section \ref{sec:model}, we introduce the \sPFC and \sAPFC models in the height formulation encoding deformations in the normal direction. For the sake of readability, the central aspects are reported in the main text, while the full expressions of quantities entering the equations and additional details are reported in the Appendices. Section \ref{sec:Model validation} is dedicated to the validation of the model. To this end, we compare special cases with established models. We start by addressing the elastic deformation of crystalline sheets under a uniaxial compressive load leading to buckling. In this situation, there are no defects. Elasticity and the different contributions entering the interplay between lattice and surface deformations are analyzed in detail, and the results obtained with the \sAPFC equations are compared with analytic solutions of the FvK equations. Next, we add defects and simulate how out-of-plane deformation of the crystalline sheet partially relaxes their long-range elastic fields. Comparisons with previous surface PFC models are drawn. In the next step, the onset of plasticity in terms of dislocation nucleation on prescribed curved but stationary surfaces is considered and compared with microscopic models. In Section \ref{sec:results}, we explore the full capability of the sAPFC model and analyze the interplay between surface out-of-plane deformation and lattice distortions. The interactions between dislocations are discussed in detail for dislocation dipoles, and the framework's capability is showcased in a simulation of a crystalline sheet hosting many dislocations. Finally, conclusions and perspectives are outlined in Section \ref{sec:conclusion}.

\section{Phase-field crystal modeling on surfaces}
\label{sec:model}

\subsection{Previous formulations}
The PFC model is based on the Swift-Hohenberg energy functional, which can be written as \citep{Elder2002,Elder2004,Emmerich2012}
\begin{equation}\label{eq:PFC_flat}
\calF_\psi=\int_{\Omega} \left[\frac{A}{2} \psi\calL\psi + \frac{B}{2}\psi^2
+\frac{C}{3}\psi^3+\frac{D}{4}\psi^4 \right]\rmd\Omega,
\end{equation}
with $\Omega \subset \mathbb{R}^2$, $\calL=(q^2+\lap)^2$, $A$, $B$, $C$, and $D$ parameters (see, e.g.,~\cite{Elder2007}), and $q$ the length of the principal (shortest) wave vector for periodic minimizers of $\calF_\psi$. In classical PFC approaches, the time evolution of $\psi\equiv\psi(\mathbf{r},t)$ with $\mathbf{r} = (x,y)$ is given by the conservative ($H^{-1}$) gradient flow of $\calF_\psi$:
\begin{equation}\label{eq:dynpfc}
    \partial_t \psi=\nabla^2 \frac{\delta \calF_\psi}{\delta \psi}.
\end{equation}
In this form, the model applies to crystals with a triangular lattice. Different dynamics and extensions have been proposed to provide advanced modeling of elastic relaxation \citep{Stefanovic2006,Toth2013,Heinonen2014,Heinonen2016,Skaugen2018,Skogvoll2022} and other lattice symmetries \citep{Wu2010,Mkhonta2013,Wang2018,Backofen2021}.

For PFC models on a curved surface, the integration over $\Omega$ needs to be replaced by an integration over the surface $\calS$ and the Cartesian Laplace operator $\nabla^2$ by the surface Laplace-Beltrami operator $\lapbel$. Such an approach has been exploited to study particle arrangements on fixed surfaces \citep{Backofen2010,Backofen2011,Kohler2016}. To account for surface deformations, a bending energy term that penalizes the surface mean curvature $\calH$ is introduced, as proposed in \cite{Aland2012} extending the PFC model. In this formulation, the free-energy functional reads
\begin{equation}\label{eq:F_PFC}
\calF_{\psi,\calS}=\int_\calS \left[\frac{A}{2} \psi\calLs\psi + \frac{B}{2}\psi^2
+\frac{C}{3}\psi^3+\frac{D}{4}\psi^4\right]\rmd\calS + \calF_b,
\end{equation}
with $\calLs=(q^2+\lapbel)^2$ and $\calF_b=\int_\calS f_b\,\rmd\calS=\int_\calS (\kappa/2)\calH^2\,\rmd\calS$ the bending energy as in the \textit{Helfrich} model \citep{Helfrich1973} with the corresponding bending energy coefficient $\kappa$. A set of equations to simultaneously relax $\psi$ and the surface $\calS$ can then be derived under the energy dissipation principle \citep{Aland2012}. Related approaches can be found in \cite{Aland2011,Aland2012b}.

The approaches mentioned above are suitable for studying crystalline arrangements on any surface. However, when considering sheets where the surface can still be described as an analytic function of a two-dimensional flat space, namely as a graph $h(x,y)$ (also referred to as \textit{Monge patches}, see \ref{app:mongepatches}), the formulation can be simplified. Such an approach has been considered to study buckling and defects in graphene and hexagonal boron nitride \citep{Elder2021,Elder2023,Granato2023}, although in a simplified setting retaining the lowest order terms only and assuming deformations along the vertical direction. As illustrated in the following sections, we consider here a height formulation that, in contrast to these previous models, accounts for deformations in the normal direction. Importantly, besides obtaining a novel \sPFC model, we devise the corresponding coarse-grained \sAPFC formulation to study elasticity directly and enable explorations at relatively large scales. 

\subsection{Surface PFC (\sPFCnospace) model with deformations in the normal direction}\label{sec:sPFC}

To study surface deformation along the surface normal, we introduce a general framework, applicable to any surface parameterization. To this end, we follow the work reported in \cite{Nitschke2020} and introduce the \textit{normal function} $\xi$, which measures the signed distance from the surface along the normal direction. We define the \textit{surface derivative} for arbitrary quantities $Q[\xi]$ depending on the normal function as
\begin{equation}
    \dS Q[\xi] \equiv \lim_{\eps \to 0} \frac{Q[\xi+\eps\delta\xi] - Q[\xi]}{\eps}.
\end{equation}
For a functional defined on the surface $\mathcal{S}$ this corresponds to
\begin{equation}\label{eq:surfelp}
    \dS Q[\xi] =\int_\mathcal{S} \frac{\delta Q}{\delta \xi} \delta \xi \,\rmd\calS,
\end{equation}
where $\rmd\calS$ is the surface element. 
Given a free energy $\calF=\calF[\pmb{X}_\xi,f]$ with $\pmb{X}_\xi$ a normal parametrization and $f$ a conserved scalar field, and assuming dissipative dynamics, we may compute the evolution via non-conservative ($L^2$) and conservative ($H^{-1}$) gradient flows for $\xi$ and $f$, respectively:
\begin{align}
    D_t{\xi} &= -M_h\frac{\delta\calF}{\delta\xi}, \\
    D_t{f} &= M_\psi\nabla_{\!\calS}^2\left(\frac{\delta\calF}{\delta f}\right), 
\end{align}
where $D_t$ indicates the material time derivative, and $M_h, \, M_\psi > 0$ are phenomenological mobility constants. For a general discussion of thermodynamically consistent gradient flows in such a setting, we refer to \cite{Nitschke2024}. The functional derivatives are weakly defined as
\begin{align}
    \scal{\frac{\delta\calF}{\delta\xi}}{\delta\xi}_E &\equiv \lim_{\eps \to 0}\frac{\calF[\pmb{X}_{\xi+\eps\delta\xi},f]-\calF[\pmb{X}_\xi,f]}{\eps} = \dS\calF,
    \\
    \scal{\frac{\delta\calF}{\delta f}}{\delta f}_E &\equiv  \lim_{\eps \to 0}\frac{\calF[\pmb{X}_\xi,f+\eps\delta f]-\calF[\pmb{X}_\xi,f]}{\eps},
\end{align}
with $E=L^2(\rmT^0\calS)$ or $H^{-1}(\rmT^0\calS)$ and $f\in\rmT^0\calS$ s.t. $\dS f=0$ holds, \ie $f$ is independent of the normal coordinate $\xi$ a priori. 

In this work, aiming at the description of crystalline sheets with relatively small deviations from flat configurations, we consider a parametrization of $\mathcal{S}$ via a height profile $h(x,y)$ with $(x,y) \in \Omega \subset \mathbb{R}^2$ (see also \ref{app:mongepatches}), while still accounting for normal deformation. Therefore, we consider the definition of the \textit{normal function} $\xi:\Omega \rightarrow\mathbb{R}$ measuring the distance from $h(x,y)$ along its normal direction (different from considering the height function that measures that distance in the vertical direction). The surface element in \eqref{eq:surfelp} reads
$\rmd\calS=\sqrt{\g}d\Omega$ with $\g=1+(\nabla h)^2$ the determinant of the metric tensor \eqref{eq:metric}. 
The \textit{normal parametrization} $\mathbf{X}_\xi=X[\xi]:\Omega\rightarrow \calS$ is then implicitly described as \citep{Nitschke2020}
\begin{equation}
    \dS\mathbf{X}[\xi] = \delta\xi \, \pmb{\nu}, 
\end{equation}
with $\pmb{\nu}$
the normal vector \eqref{eq:appSurfaceNormal}. Useful geometrical quantities and identities are reported in \ref{app:mongepatches}. Then, using the frame independence of the normal velocity \citep{Nitschke2020}, we have:
\begin{equation}
    D_t\xi=v_\perp=\scal{\dt\mathbf{X}}{\pmb{\nu}}=\scal{(\dt h)\mathbf{e}_z}{\pmb{\nu}}\stackrel{\eqref{eq:appSurfaceNormal}}{=}\frac{\dt h}{\sqrt{\g}}.
\end{equation}
with $D_t$ the material time derivative. In general, applied to a scalar field $f$, it reads $D_t f = \dt f+\nabla_{\!\mathbf{v}} f$,
where $\mathbf{v}$ is the \textit{relative velocity} given by $\mathbf{v}=\mathbf{V}_\xi-\mathbf{V}$, with $\mathbf{V}_\xi=D_t\xi\,\pmb{\nu}$ the \textit{material velocity} and $\mathbf{V}=(\dt h)\mathbf{e}_z$ the \textit{observer velocity}. In the height formulation, the relative velocity may then be cast as
\begin{equation}
\begin{split}
    \mathbf{v} =& \frac{\dt h}{\sqrt{\g}}(\mathbf{e}_z - \delta^{ij}(\partial_i h)\mathbf{e}_j) - (\dt h)\mathbf{e}_z
    \\
    =& -(\dt h)
    \frac{\delta^{ij}(\partial_i h)}{\sqrt{\g}}
    \underbrace{(
    \mathbf{e}_j +(\partial_j h)\mathbf{e}_z)}_{\displaystyle=\partial_j\pmb{X}} = -(\dt h)\nablas h,
\end{split}
\end{equation}
and thus, the time derivative reads
\begin{equation}
    D_t{f} = \dt f - (\dt h)\frac{\scal{\nablaf h}{\nablaf f}}{\g}.
\end{equation}
With these notions, we can write the \sPFC model in the height formulation encoding deformation along the surface normal as
\begin{align}
    \dt h &= -M_h\sqrt{\g}\left(\frac{\delta\calF_{\cal{L}}}{\delta\xi}+\frac{\delta\calF_{b}}{\delta\xi}\right), 
    \label{eq:gradient-flow_1}
    \\
    \dt\psi &= (\dt h)\frac{\scal{\nablaf h}{\nablaf\psi}}{\g} + M_\psi\nabla_{\!\calS}^2\left(\frac{\delta\calF_{\psi,\calS}}{\delta\psi}\right), 
    \label{eq:gradient-flow_2}
\end{align}
with $\calF_{\cal{L}}=\int_\calS [\frac{A}{2} \psi\calLs\psi] \rmd\calS$.
We remark that, although we consider a graph formulation and describe the surface evolution via $\partial_t h$, this quantity is determined by variation of $\calF_{\cal{L}}$ and $\calF_{b}$ (the bending energy) with respect to $\xi$, encoding the effects of deformation along the direction normal to the surface. 
Additionally, note that as the embedding space has implicitly zero energy, considering the variation of the whole functional $\calF_{\psi,\cal{S}}$ would lead to out-of-plane growth of crystalline sheets for bulk energies different than zero, which is a different physical setting from the one we are interested in (for instance growth against a surrounding liquid phase). Consequently, we retain only the variation of $\calF_{\cal{L}}$ and $\calF_{b}$, which corresponds to inspecting out-of-plane deformations due to elasticity effects exclusively \citep{Salvalaglio2022ov}.  Similar assumptions entered previous PFC formulations accounting for out-of-plane deformations  \citep{Elder2021}.
The explicit expression for the variational derivatives entering Eqs.~\eqref{eq:gradient-flow_1}--\eqref{eq:gradient-flow_2} are reported in \ref{app:FuncDer}. To numerically solve these equations, we consider the numerical method outlined in \ref{app:numerics}.

We finally note that in the limit of infinitely small slopes, $\xi \sim  h$. By neglecting higher order terms in derivatives of $h$, Eqs.~\eqref{eq:gradient-flow_1}--\eqref{eq:gradient-flow_2} indeed reduce to previously proposed PFC formulations inspecting deformation along the vertical direction only \citep{Elder2021,Elder2023,Granato2023}. Different approximations are further discussed below in Sect.~\ref{sec:Model validation}. Finally, in Sect.~\ref{sec:results}, we examine applications where the small slopes approximation does not hold, motivating the more accurate description of surface deformation along its normal.

\subsection{Surface APFC (sAPFC) model with deformations in the normal direction}
\label{sec:sAPFC}

The coarse-graining achieved by the APFC model is based on the approximation of $\psi$ by a sum of plane waves with complex amplitudes $\eta_j$ \citep{Goldenfeld2005,Athreya2006,Salvalaglio2022ov}: 
\begin{equation}
    \psi(\mathbf{r}) = \no + \sum_{n=1}^N \eta_n e^{\i\kv_n\cdot \mathbf{r}}+\text{c.c.}=
    \sum_{n=-N}^N  \eta_n e^{\i\kv_n\cdot \mathbf{r}},
    \label{eq:n_app}
\end{equation}
where $\eta_{-n}=\eta^*_n$, $\eta_0=\no$, $\kv_n$ are the reciprocal space vectors reproducing a specific lattice symmetry, with $\kv_{-n}=-\kv_n$, $\kv_0=\mathbf{0}$, $\i$ the imaginary unit, and c.c. referring to the complex conjugate. Usually, a small set of modes is considered, with ``mode" referring to a family of equal-length $\kv_n$.
A free energy functional $\calF_\eta$ depending on $\eta_n$ can be formally obtained through a renormalization group approach \citep{Goldenfeld2005} or, equivalently, by substituting Eq.~\eqref{eq:n_app} into Eq.~\eqref{eq:PFC_flat} and integrating over the unit cell under the assumption of constant amplitudes therein \citep{Athreya2006,Salvalaglio2022ov}. By following this derivation and accounting for the integration over the surface $\mathcal{S}$ as in Eq.~\eqref{eq:F_PFC}, we get
\begin{equation}
    \calF_{\eta,\calS} = \int_\calS \left[A\sum_{n=1}^N\left|\calG^\calS_n\eta_n\right|^2 + g^s(\{\eta_n\},\overline{\psi}) \right]\rmd\calS + \mathcal{F}_b,
\label{eq:F_APFC} 
\end{equation}
where
\begin{equation*}
\begin{split}
    \calG^\calS_n\eta_n =& \lapbel\eta_n + (q^2-||\kv_n||_\calS^2)\eta_n + \i\left(2\scal{\kv_n}{\nablas\eta_n}_\calS + \mathrm{div}_\calS(\kv_n)\eta_n\right), \\
    g^s =& \frac{B+2C\no+3D\no^2}{2}\zeta_2 + \frac{C+3D\no}{3}\zeta_3 + \frac{D}{4}\zeta_4 + \frac{A}{2}\no^2 \\
    &+\frac{C}{3}\no^3+\frac{D}{4}\no^4, \\
    \zeta_2 =& \sum_{(p,q)\in\{-N..N\}^2}\eta_p\eta_q \delta_{\mathbf{0},\kv_p+\kv_q}
    =2\sum_{n=1}^N |\eta_n|^2=\Phi,
    \\
    \zeta_3 =& \sum_{(p,q,r)\in\{-N..N\}^3}\eta_p\eta_q\eta_r \delta_{\mathbf{0},\kv_p+\kv_q+{\kv_r}},
    \\
    \zeta_4 =& \sum_{(p,q,r,s)\in\{-N..N\}^4}\eta_p\eta_q\eta_r\eta_s \delta_{\mathbf{0},\kv_p+\kv_q+{\kv_r}+\kv_s},
    \end{split}
\end{equation*}
with $\delta_{\mathbf{0},\mathbf{Q}}=1$ if $\mathbf{Q}=\mathbf{0}$ and 0 otherwise. See \ref{app:mongepatches} for explicit definitions of $||\,\cdot\,||_\calS$, $\nabla_\calS$, and ${\rm div}_\calS$ and other related quantities. To quantify the energetics of deformed sheets, we will consider the \textit{deformation energy} defined as $\calF_{\eta,\calS}-\calF_0$ with $\calF_0$ the total energy in a relaxed, flat sheet. We remark that the considered formulation of the \sAPFC model implies that the scalar product $\mathbf{k}_n\cdot \mathbf{r}$ entering Eq.~\eqref{eq:n_app} (and $\mathcal{G}_n$) defined as in the flat space is a good approximation of the inner product on the tangent space. The latter could be exactly represented by considering a geodesic distance but would result in an extremely expensive method from a computational point of view (involving an additional differential problem to solve per discretization point). We will explicitly show that this convenient approximation works very well for the applications discussed in this work by comparing the results of the \sAPFC model with the \sPFC model, where this approximation is not needed (see Section \ref{sec:dislonucleation} and \ref{app:validation}). Explicit forms of $g^{\rm s}$ for different lattice symmetries can be found in \cite{Salvalaglio2022ov}. In this work, we consider the triangular and the hexagonal lattices, the latter corresponding to the technology-relevant cases of graphene \citep{Hirvonen2017} or hexagonal boron nitride \citep{Molaei2021} sheets. They can be simulated by considering the one-mode approximation of the triangular lattice (with the hexagonal structure being its dual lattice) in a certain parameter range, setting $\mathbf{k}_1=[-\sqrt{3}/2,-1/2]$, $\mathbf{k}_2=[0,1]$, $\mathbf{k}_3=[\sqrt{3}/2,-1/2]$. This choice results in $\zeta_3=\eta_1\eta_2\eta_3+\eta_1^*\eta_2^*\eta_3^*$ and $\zeta_4=0$. The description of hexagonal lattices naturally follows from considering a triangular lattice with a proper basis as discussed in \cite{DeDonnoPRM2023} and has been proven capable of describing general properties and defects therein \citep{Mkhonta2013,Hirvonen2017}. 

The classical dynamics of the APFC model (\ie for a flat surface) is given by an approximation of Eq.~\eqref{eq:dynpfc} reading $\partial_t\eta_n=-|\mathbf{k}_n|^2\delta F_\eta / \delta \eta_n^*$ \citep{Salvalaglio2022ov} and a conservative ($H^{-1}$) gradient flow to compute $\partial_t \overline{\psi}$ \citep{Yeon2010}.
Following the same arguments as in the previous sections, which lead to expressions in the height formulation of the gradient flows, we finally obtain the dynamical equations of the \sAPFC model:
\begin{align}
\dt h =& -M_h\sqrt{\g}\left(\frac{\delta\calF_{\calG}}{\delta\xi}+\frac{\delta\calF_{b}}{\delta\xi}\right), \label{eq:hapfc}
\\
\dt\eta_n =& (\dt h)\frac{\scal{\nablaf h}{\nablaf\eta_n}}{\g} 
\label{eq:etaapfc}
\\
&-M_\eta\left(||\kv_n||^2-\frac{\scal{\kv_n}{\nablaf h}^2}{\g}\right)\left(A\left(\calG_n^\calS\right)^2\!\!\eta_n +\frac{\partial g^s}{\partial\eta_n^*}\right),\nonumber
\\
\dt\no =& (\dt h)\frac{\scal{\nablaf h}{\nablaf\eta_n}}{\g} + M_\no\lapbel\left(\frac{\delta\calF_{\eta,\calS}}{\delta\no}\right), \label{eq:barpsi}
\end{align}
with $\calF_\calG=A\sum_{n=-N}^N\int_\calS\left|\calG_n^\calS\eta_n\right|^2\rmd\calS$ the elastic energy, whose variation w.r.t. $\xi$ determines the out-of-plane deformations together with the variation of $\calF_b$, similarly to the \sPFC model introduced in Section \ref{sec:sPFC}. $M_h, M_\eta$, and $M_{\bar \psi} > 0$ are phenomenological mobility constants. Functional derivatives and other quantities entering Eqs.~\eqref{eq:hapfc}--\eqref{eq:barpsi} are reported in \ref{app:FuncDer}. To numerically solve these equations, we consider the numerical method outlined in \ref{app:numerics}. 

The APFC model is often solved in a slightly simplified setting that neglects Eq.~\eqref{eq:barpsi}, \ie assuming a constant $\overline{\psi}$. This is a good approximation for bulk systems, although it leads to some differences between the PFC and APFC formulations, particularly for phase transitions. For the applications of interest in this work, namely sheets of a stable crystalline phase, this approximation is considered. However, for completeness, a dedicated discussion showcasing the possibility of considering the full \sAPFC model is reported in \ref{app:DAPFC}.

\subsection{F\"oppl-von K\'arm\'an equations}
To validate our model with existing continuum models, we also introduce the F\"oppl-von K\'arm\'an (FvK) equations. Specified to our case of interest, the equations reduce to:
\begin{equation}\label{eq:fvk}
\left\{
\begin{aligned}
    &\frac{1}{E}\nabla^4\varphi + \frac{1}{2}[h,h] = 2e_{\alpha\beta}(\partial_\beta J^\alpha), \\
    &\kappa\nabla^4h - [h,\varphi] = 0,
\end{aligned}
\right.
\end{equation}
where $\varphi$ denotes the Airy stress function, $E$ the two-dimensional stretching modulus ($E=Yd$ with $Y$ the Young modulus and $d$ the surface thickness), $e_{\alpha\beta}$ the permutation symbol, $J^\alpha$ the density of dislocations with Burgers vector along the $\alpha$-direction, and $[\cdot,\cdot]$ the two-dimensional Monge–Ampère bracket defined by $[f,g]=(\partial_{xx}f)(\partial_{yy}g) + (\partial_{yy}f)(\partial_{xx}g) - 2(\partial_{xy}f)(\partial_{xy}g)$. Without dislocations, Eqs. \eqref{eq:fvk} reduce to the classical FvK equations \citep{Seung1988,Zhang2014a}, while a generalized version of the FvK equations with defects is given in \cite{Singh2022}.

We also report the associated energy, which allows for a clearer comparative analysis with the proposed sPFC and sAPFC models:
\begin{equation}\label{eq:fvk_energy}
\calF_\text{FvK} = \frac{1}{2}\int_\Omega \left[\kappa(\nabla^2 h)^2 + \lambda U_{ii}^2 + 2\mu U_{ij}^2 \right] \,\rmd\Omega,
\end{equation}
with $U_{ij} = \left[\partial_i u_j + \partial_j u_i + (\partial_i h)(\partial_j h)\right]/2$, $\mathbf{u} = (u_x,u_y)$ the displacement field, and $\lambda$, $\mu$ denote the two-dimensional Lam\'e parameters, such that $E=4\mu(\lambda+\mu)/(\lambda+2\mu)$.

\section{Model validation} 
\label{sec:Model validation}

We consider various subproblems with increasing complexity to validate the proposed \sAPFC model. Each subproblem is compared with analytical results, continuous modeling approaches, or microscopic models. Further details, also addressing a comparison of \sPFC and \sAPFC models, are outlined in \ref{app:validation}.

\subsection{Buckling under a compressive load}
\label{sec:buckling}
\begin{figure}
    \centering
    \includegraphics[width=0.5\textwidth]{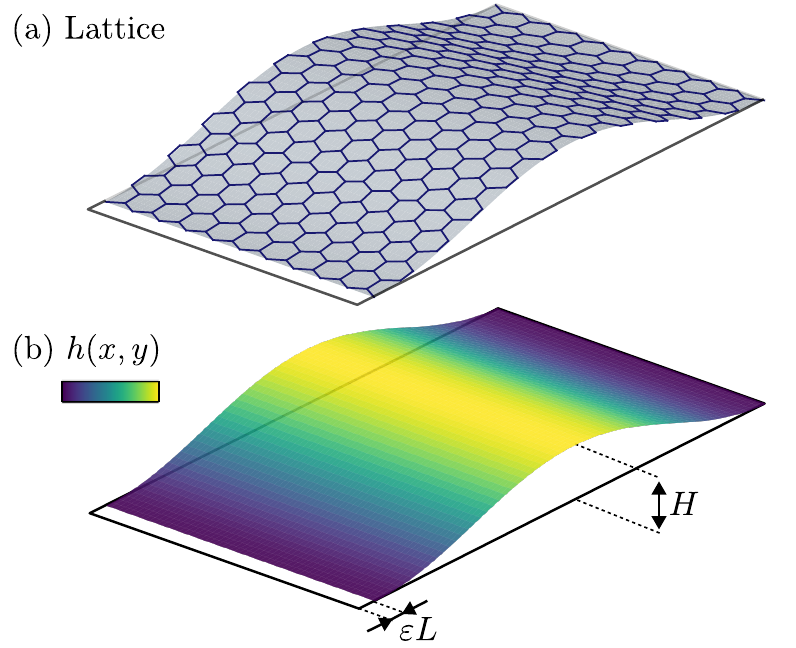}
    \caption{Configuration for analyzing the buckling of a crystalline sheet. (a) Perspective view of the reconstructed lattice structure, illustrated as the hexagonal arrangement connecting the maxima of the (reconstructed) microscopic density field $\psi$ with their nearest neighbors. (b) Height profile with parametrization considered in Section~\ref{sec:buckling} obtained by an sAPFC simulation. The system is four time larger than panel (a), featuring a strain compatible with elastic regimes in crystalline systems. The black lines represent the flat configuration. $H$ denotes the maximum height, $L$ the length of the sheet and $\eps$ the magnitude of the uniaxial strain.}
    \label{fig:buckling_schematic}
\end{figure}

\begin{figure*}
\centering
    \includegraphics[width=0.98\textwidth]{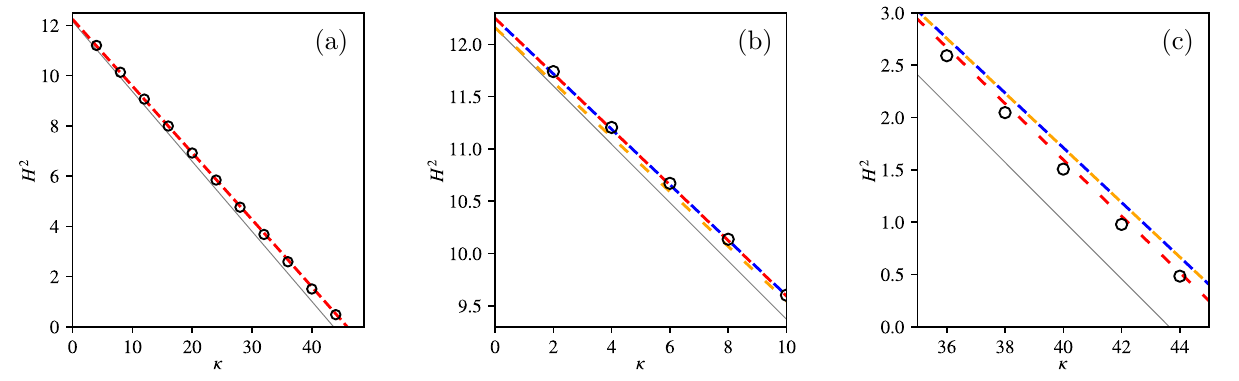}
    \caption{Buckling of a crystalline sheet in the elastic regime. (a) \sAPFC simulation results (black circles) in terms of $H^2$ (see parametrization in Fig.~\ref{fig:buckling_schematic}) against bending stiffness $\kappa$.
     We consider a system with $N=40$ the number of atoms in the compression direction, and uniaxial strain $\varepsilon=-1/40=-0.025$, and use the following set of parameters: $(A, B, C, D) = (1, -0.15, 1/2/\sqrt{0.98/3}, 1)$ \citep{Hirvonen2017}, which leads to a honeycomb lattice structure. Simulation results are compared with different analytical approximations (lines): gray $\bigO(\eps_0^2, g^3)$, orange $\bigO(\eps_0^3, g^3)$, blue $\bigO(\eps_0^3, g^5)$, and red $\bigO(\eps_0^4, g^5)$. (b) and (c) show details of the plot in (a) at low and large bending stiffnesses $\kappa$, respectively.}
    \label{fig:buckling_analytic}
\end{figure*}

To showcase the description of the elastic behavior encoded in the \sAPFC model, we start by examining a defect-free crystalline sheet under a uniaxial compressive load. Indeed, in this simplified setting, an analytic solution is available for the FvK model, and approximate analytic expressions may be derived for the sAPFC model.

As illustrated in Fig.~\ref{fig:buckling_schematic}, we consider a uniform sheet of length $L$ in the direction of compression, with a uniaxial strain of magnitude $\eps < 0$, which buckles with a maximum height $H \ge 0$. In the (s)APFC model, the displacement field $\mathbf{u}$ is encoded in the phase of the complex amplitudes \citep{Salvalaglio2022ov}:
\begin{equation}
    \eta_n = \phi_n e^{-\i\mathbf{k}_n\cdot\mathbf{u}}, \quad \phi_n\in\mathbb{R}.
    \label{eq:ampdisp}
\end{equation}
Introducing this expression in the energy \eqref{eq:F_APFC}, the stationary solution is given by the conditions:
\begin{equation}
    \frac{\delta \calF_\eta}{\delta \phi_n} = 0, \quad \frac{\delta \calF_\eta}{\delta h} = 0, \quad \frac{\delta \calF_\eta}{\delta \mathbf{u}} = \mathbf{0}.
\end{equation}
These equations yield a highly nonlinear system of coupled equations that cannot be easily solved in its current form. We proceed further via Taylor expansions near the critical buckling point, corresponding to where the sheet is maximally compressed while remaining flat. Importantly, we take advantage of two different approximations: a \textit{geometric one} (small surface height gradients $HQ \ll 1$) and a \textit{physical one} (small deformations $|\eps| \ll 1$), where $Q=2\pi/L$ denotes the wave number of the buckling profile. At the critical buckling point, assuming without loss of generality that the sheet is compressed along the $y$-axis, the height profile and displacement field take the form:
\begin{equation}
    h_0(y) = 0, \quad \mathbf{u}_0(x,y) = u_0(y)\mathbf{e}_y = \epsc y\, \mathbf{e}_y.
\end{equation}
In the limit of small surface height gradients, \ie $g \equiv H Q \ll 1$, they can be expanded such that:
\begin{align*}
    h(y) &= h_0(y) + g h_1(y) + g^2 h_2(y) + \bigO(g^3), \\
    u(y) &= u_0(y) + g u_1(y) + g^2 u_2(y) + \bigO(g^3).
\end{align*}
At lowest order, the condition $\delta\calF_\eta/\delta h = 0$ yields:
\begin{equation*}
h_1(y) = H \cos\left((1-\epsc)\sqrt{\frac{A}{4\kappa}|\epsc|(2-\epsc)(\phi_1^2+16\phi_2^2+\phi_3^2)}\,y\right),
\end{equation*}
so that, by definition,
\begin{equation}\label{eq:epsc}
    Q = (1-\epsc)\sqrt{\frac{A}{4\kappa}|\epsc|(2-\epsc)(\phi_1^2+16\phi_2^2+\phi_3^2)}.
\end{equation}

\begin{figure*}
    \centering
    \includegraphics[width=\textwidth]{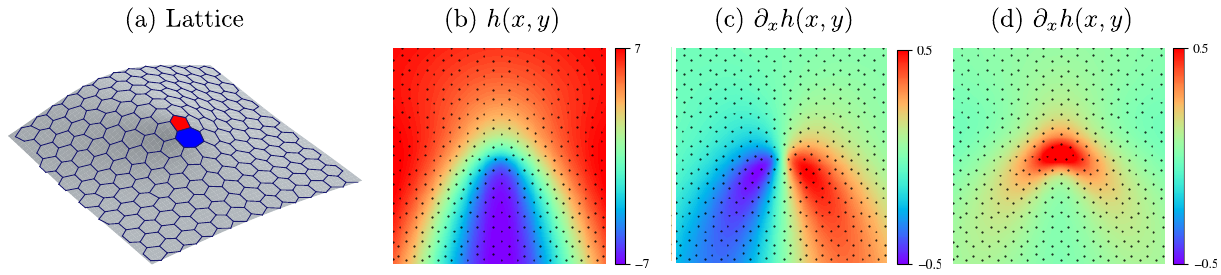}
    \caption{Out-of-plane deformation at a dislocation by the \sAPFC model. A small portion (13$a_x$ $\times$ 13$a_x$) centered around a dislocation of a domain of total size (141$a_x$ $\times$ 40$a_y$) is shown for $\kappa = \kappa_g \simeq 0.21$, chosen as in \citep{Dai2016} to match properties of graphene, where $a_x=4\pi/\sqrt{3}$ and $a_y=4\pi$ are the lattice parameters in the $x$ and $y$ directions, respectively. (a) Perspective view of the reconstructed lattice structure hosting a $5|7$ defect as in Fig.~\ref{fig:buckling_analytic}(a). (b) surface height profile $h$, (c) $\partial_x h$, and (d) $\partial_y h$. The colormap was chosen to match the one in \cite{Elder2021}. The maxima of the reconstructed density are shown as black dots in panels (b)--(d). Energy parameters are set as in Section~\ref{sec:buckling}: $A=1, B=-0.15, C=0.87$, and $D=1$.}
    \label{fig:disloc_profile}
\end{figure*}

With the condition $\delta\calF/\delta\eta_n=0$, for a hexagonal lattice, we have
\begin{equation}\label{eq:amp_expansion}
\left\{
\begin{aligned}
\phi_n &= \phi_{n,0} + \epsc^2 \phi_{n,2} + \bigO(\epsc)^3, \\
\phi_{1,0} &= \phi_{2,0} = -\phi_{3,0} = \phi = \frac{C+\sqrt{C^2-15BD}}{15D}, \\
\phi_{1,2} &= -\phi_{3,2} = \frac{A}{8}\times\frac{136C^2+93BD-675CD\phi}{(16C^2+3BD)(C-15D\phi)}, \\
\phi_{2,2} &= \frac{A}{4}\times\frac{8C^2-66BD+675CD\phi}{(16C^2+3BD)(C-15D\phi)}, 
\end{aligned}\right.    
\end{equation}
and thus, inserting \eqref{eq:amp_expansion} into \eqref{eq:epsc}:
\begin{equation}
    \epsc = \eps_0 + \frac{5}{2}\eps_0^2 + \bigO(\eps_0^3), \quad \eps_0 = -\frac{\kappa Q^2}{9A\phi^2}.
\end{equation}
Similarly, we compute:
\begin{equation}
h_2(y) = u_1(y) = 0, \quad u_2(y) = \eps_2 y + U_2\sin(2Qy), 
\end{equation}
with
\begin{equation}
\left\{
\begin{aligned}
    \eps_2 &= -\frac{1}{4} - \frac{15\eps_0-8Q^2}{24} + \bigO(\eps_0^2), \\
    Q U_2 &= \frac{1}{8} - \frac{3\eps_0-4Q^2}{12} + \bigO(\eps_0^2).
\end{aligned}
\right.
\end{equation}
From the boundary condition
\begin{equation}
    u(L) = \epsc L + (HQ)^2\eps_2 L = \eps L,
\end{equation}
we conclude, assuming $Q \sim \eps$:
\begin{equation}\label{eq:hq}
    (HQ)^2 = \frac{\eps-\epsc}{\eps_2} = -4(\eps-\eps_0) + 10 \eps_0\eps + \bigO(\eps_0^3).
\end{equation}
The derivation for higher-order approximations follows the same procedure, and we only report the final expression for brevity. Taking into account the next relevant orders in surface height gradients $\bigO(g^5)$ and critical strain $\bigO(\eps_0^4)$, we obtain:
\begin{align}\label{eq:highorder}
\begin{split}
    (HQ)^2 = &-4(\eps-\eps_0) + \frac{7}{6}\left(\eps^2+\eps_0^2+\frac{46}{7}\eps_0\eps\right) \\
    &- \frac{49}{72}\eps^3 + \frac{19}{2}\eps^2\eps_0 + \frac{245}{24}\eps_0^2\eps \\
    &- \frac{73}{36}\eps_0^3\left(1+\frac{32}{73}\frac{\phi_1 + 8\phi_2}{\phi}\right)+ \bigO(\eps_0^4).
\end{split}
\end{align}

To test this prediction, we solve the sAPFC model in this setting for different bending stiffnesses $\kappa$ and a compressive uniaxial strain $\varepsilon_0=-1/40=-0.025$. This is achieved by setting the amplitudes as described in \eqref{eq:ampdisp}, with the (linear) displacement $(u_x, u_y)=(0, \varepsilon_0 y)$. Results are reported in Fig.~\ref{fig:buckling_analytic}. The lowest order approximation (gray solid line) captures the overall correct behavior, with some deviations for small and large bending stiffness. By considering a higher order of surface height gradient ($g$), excellent agreement is now obtained in the region of low bending stiffness/large buckling height. Similarly, higher orders of the critical strain ($\varepsilon_0$) lead to excellent agreement in the region of large bending stiffness/small buckling height. This allows for a direct characterization of higher order \textit{geometrical} and \textit{physical} terms encoded in the sAPFC model.

For comparison, we also derive the corresponding relation from the standard FvK theory as reported in Eqs.~\eqref{eq:fvk}, with $\mathbf{J}=0$. In the one-dimensional setting, Eqs. \eqref{eq:fvk} reduce to:
\begin{equation}
\left\{
\begin{aligned}
0 &= \frac{\rmd}{\rmd y}\left[\frac{\rmd u}{\rmd y} + \frac{1}{2}\left(\frac{\rmd h}{\rmd y}\right)^2\right], \\
0 &= \kappa \frac{\rmd^4 h}{\rmd y^4} - (\lambda+2\mu)\frac{\rmd}{\rmd y}\left[\left(\frac{\rmd u}{\rmd y} + \frac{1}{2}\left(\frac{\rmd h}{\rmd y}\right)^2\right)\frac{\rmd h}{\rmd y}\right],
\end{aligned}
\right.
\end{equation}
For the first bending mode, this yields:
\begin{equation}
h = \frac{2}{Q}\sqrt{|\eps| - \frac{\kappa Q^2}{\lambda+2\mu}} \cos(Q x)
\end{equation}
and thus
\begin{equation}
    (HQ)^2 = -4(\eps-\eps_0),
\end{equation}
under the identity $\lambda=\mu=3A\phi^2$ for the triangular lattice in the APFC model. Consequently, we identify the standard FvK theory as the lowest-order approximation of the \sAPFC model. To further illustrate this point, we recall the expression for the free energy of the crystalline sheet within the FvK framework \eqref{eq:fvk_energy} which relies on a linear stress-strain relation and a linearized bending force, explaining the differences with the \sAPFC predictions observed in Fig.~\ref{fig:buckling_analytic} at large and low bending stiffnesses $\kappa$, respectively. Indeed, in the region of large bending stiffness, the in-plane strain is high enough that the nonlinear elastic contributions of the \sAPFC model become significant. Likewise, at low bending stiffness, the buckling height is such that higher-order height-gradient terms start playing a non-negligible role.

\subsection{Surface profile at a dislocation}
\label{sec:surfdislo}

We now introduce dislocations and analyze the out-of-plane deformation they induce. Dislocations generate long-range elastic fields, which can be reduced by out-of-plane displacement. They can be initialized by imposing the corresponding displacement field (e.g. from \cite{anderson2017}) into Eq.~\eqref{eq:ampdisp}, see also \cite{Salvalaglio2022ov}. In particular, we analyze the relaxation of a sheet hosting four dislocations arranged in two pairs. While on a flat-constrained sheet, this defect configuration is stable due to the symmetries of the system, it is no longer the case when the fully coupled model is considered, and the out-of-plane displacement breaks the symmetry: the dislocations are observed to migrate away from the bulge they create (see further discussion on this aspect in Section \ref{sec:interaction}). In the \sPFC model, this behavior should be expected close to the melting point only as it involves a motion by climbing that is prevented by pinning at relatively low quenching depths \citep{Skaugen2018a,Elder2021}. To obtain a pseudo-stationary surface height profile, we reinitialize the position of the considered four defects in their square subdomains (squares centered in $(\pm L/2,\pm L/2)$ with $L$ the side of the simulation domain) at every step. This is formally done by translating the solutions of the dynamical equations as   
    $\tilde{h}(\mathbf{x}) = h(\mathbf{x}-\mathcal{T}*\bfdelta_i)$ and $\tilde{\eta}_n(\mathbf{x}) = \eta_n(\mathbf{x}-\mathcal{T}*\bfdelta_i)$
with $\bfdelta_i$ the dislocation shift of the i-th dislocation with respect to their initial positions, and $\mathcal{T}$ a smoothing kernel avoiding sharp transitions at the subdomains' borders. This procedure converges numerically.

A portion of the system near one of the dislocation cores is presented in Fig.~\ref{fig:disloc_profile}. The resulting height profile and its gradients are in very good agreement with the ones considered in \cite{Singh2022} for the FvK model with defects, as well as experimental \citep{Lehtinen2013} and atomistic \citep{Zhang2014a} results. The results also resemble those obtained by the surface PFC model \citep{Elder2021}. The surface is essentially flat on the pentagon (tensile) side of the dislocation core but shows significant variations on the heptagon (compressive) side, see Fig.~\ref{fig:disloc_profile}(b). Note that although the \sAPFC model solves for slowly-varying amplitudes, it is straightforward to reconstruct the corresponding atomic density from them via Eq.~\eqref{eq:n_app} and recover the typical 5|7 defect. An illustration is shown in Fig.~\ref{fig:disloc_profile}(a).

\begin{figure}[t!]
    \centering
    \includegraphics[width=0.95\linewidth]{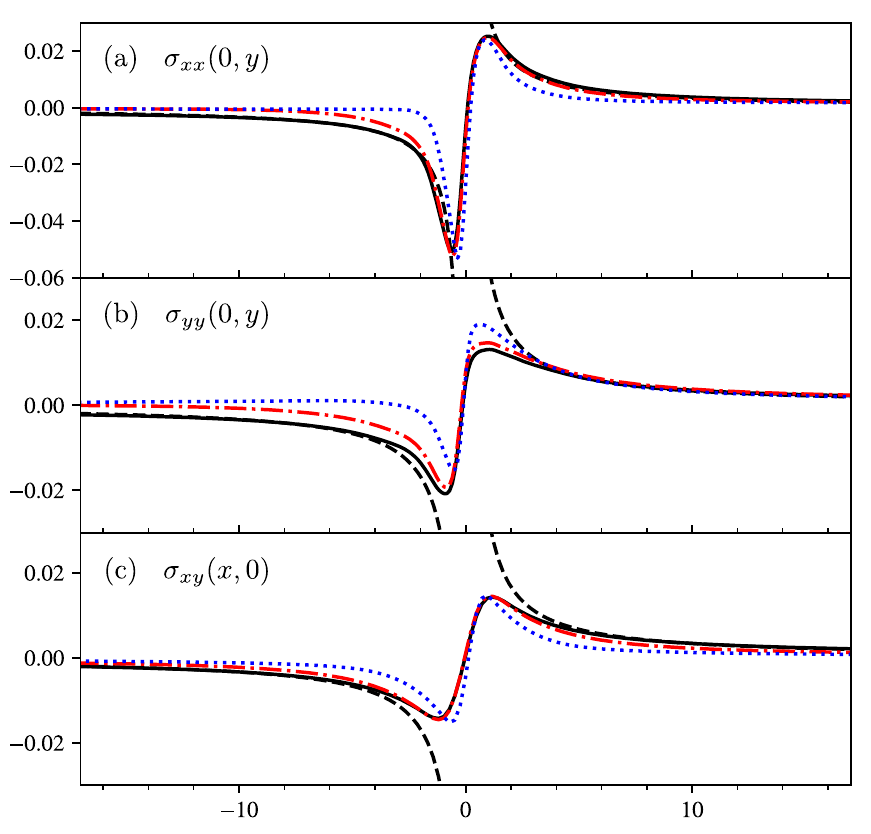}
    \caption{Stress field at a dislocation computed by equation \eqref{eq:stress} from the results of \sAPFC simulations with: $\kappa = \kappa_g \simeq 0.21$ (blue dotted line), $\kappa = 4\kappa_g \simeq 0.84$ (red dash-dotted line), rigid sheet $\kappa \rightarrow \infty$ (black solid line). For completness, the prediction of continuum mechanics in a flat domain \citep{anderson2017} is also shown (black dashed line). Energy parameters are set as in Fig.~\ref{fig:disloc_profile}.}
    \label{fig:disloc_stress}
\end{figure}

To study the influence of the bending stiffness $\kappa$, the stress field has been calculated from the amplitudes extending the approaches proposed in \cite{Skaugen2018a,Salvalaglio2019,Salvalaglio2020plastic} to deformable surfaces, which yields:
\begin{equation}
\begin{split}
    \frac{\sigma_{\alpha\beta}}{A} = &-\sum_{n=-N}^N \Bigg\{ v_\alpha \left[\calG_n^\calS\eta_n\right]\left[(\partial_\beta-\i k_{n,\beta})\eta_n^*\right] 
    \\
    &
    + \sum_{\gamma=1}^2 \Bigg(  w_{\alpha \gamma}  \left[(\partial_\gamma + \i k_{n,\gamma})\calG_n^\calS\eta_n\right]\left[(\partial_\beta-\i k_{n,\beta})\eta_n^*\right]
    \\
    &- \left[\calG_n^\calS\eta_n\right]\left[(\partial_\beta-\i k_{n,\beta})(\partial_\gamma-\i k_{n,\gamma})\eta_n^*\right] 
    \\
    &
    + (\partial_\gamma w_{\alpha\gamma})\left[\calG_n^\calS\eta_n\right]\left[(\partial_\beta-\i k_{n,\beta})\eta_n^*\right]\Bigg) \Bigg\}
    \label{eq:stress}
\end{split}
\end{equation}
where
\begin{align}
    v_\alpha &= \frac{\partial(\nablaf^2_\calS \psi)}{\partial(\partial_\alpha\psi)} = \frac{-\calH}{\sqrt{\g}}\,\partial_\alpha h,
    \\
    w_{\alpha\gamma} &= \frac{\partial(\nablaf^2_\calS \psi)}{\partial(\partial_{\alpha\gamma}\psi)} = \delta_{\alpha\gamma} - \frac{(\partial_\alpha h)(\partial_\gamma h)}{\g}.
\end{align}

In the flat case, $\pmb{v}=\pmb{0}$ and $\pmb{w}=\mathbf{Id}$, thus recovering the formula for a flat surface. Note that in microscopic/atomistic methods, a coarse-graining step (spatially and/or temporally) is required to infer the stress field from microscopic quantities. By contrast, as APFC models solve for the amplitudes directly, they grant straightforward access to the stress field at the continuum level.

The calculated stress fields are detailed in Fig.~\ref{fig:disloc_stress}. We observe a regularization of the stress field at the dislocation core, as is typical for (A)PFC models \citep{Skogvoll2023,LBM_GE_2024gradient}. More importantly, we note that out-of-plane displacements lead to the localization of the compressive stress closer to the dislocation core. This is a result of buckling. Comparing different values for the bending stiffness $\kappa$, shows that softer crystalline sheets (smaller $\kappa$) allow for increased localization. 

\subsection{Defect arrangements on a fixed surface profile}
\label{sec:dislonucleation}

\begin{figure}[b!]
    \centering
    \includegraphics[width=0.95\linewidth]{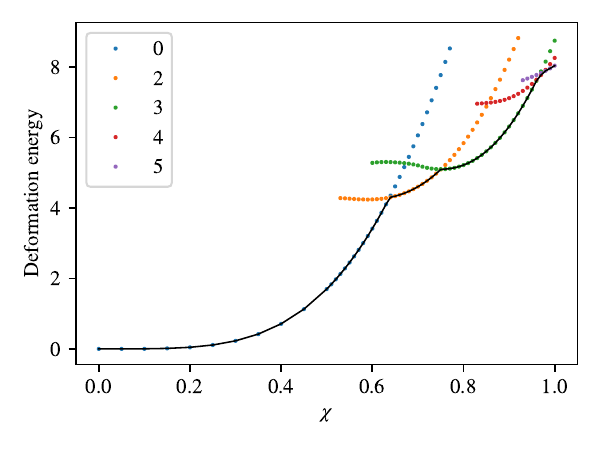}
    \caption{Deformation energy $\calF_{\eta,\calS}-\calF_0$ as a function of the aspect ratio $\chi$ for the \sAPFC model. The dotted curves correspond to simulations with a given number of dislocations (see legend). Energy parameters (see Eq.~\eqref{eq:F_APFC}): $A=2$, $B=0.02$, $C=-1/2$, and $D=1/3$.}
    \label{fig:bump_apfc}
\end{figure}
We now examine the nucleation, stability, and arrangement of defects on a fixed surface profile. In other words, while the general \sAPFC model realizes a two-way coupling between the amplitude fields and the surface height profile, we here restrain it to a one-way coupling whereby the amplitudes adjust to the surface profile without affecting it (corresponding to $M_h=0$ in Eq. \eqref{eq:hapfc}).  
Note that this phenomenology, as well as the tracking of defect migration and annihilation further explored in the following, is typically not accessible within continuum approaches without prescribing the defect distribution and ad-hoc rules. In this section, comparisons will be drawn with atomistic-scale Monte Carlo (MC) simulations \citep{Hexemer2007}.

To analyze the interplay between out-of-plane deformations and lattice distortions, and specifically assess the influence of curvature on defect configurations, we consider the shape of a Gaussian bump with standard width $\rho$ and varying aspect ratio $\chi$ such that
\begin{equation}
    h(x,y) = \chi \rho e^{-\frac{x^2+y^2}{2\rho^2}}.
    \label{eq:gprof}
\end{equation}
Indeed, out-of-plane deformations stretch the lattice, creating strains that may be relieved by an appropriate distribution of defects, depending on the competition between the elastic strain energy and the defect energy.

\begin{figure*}
    \centering
    \includegraphics[width=0.73\linewidth]{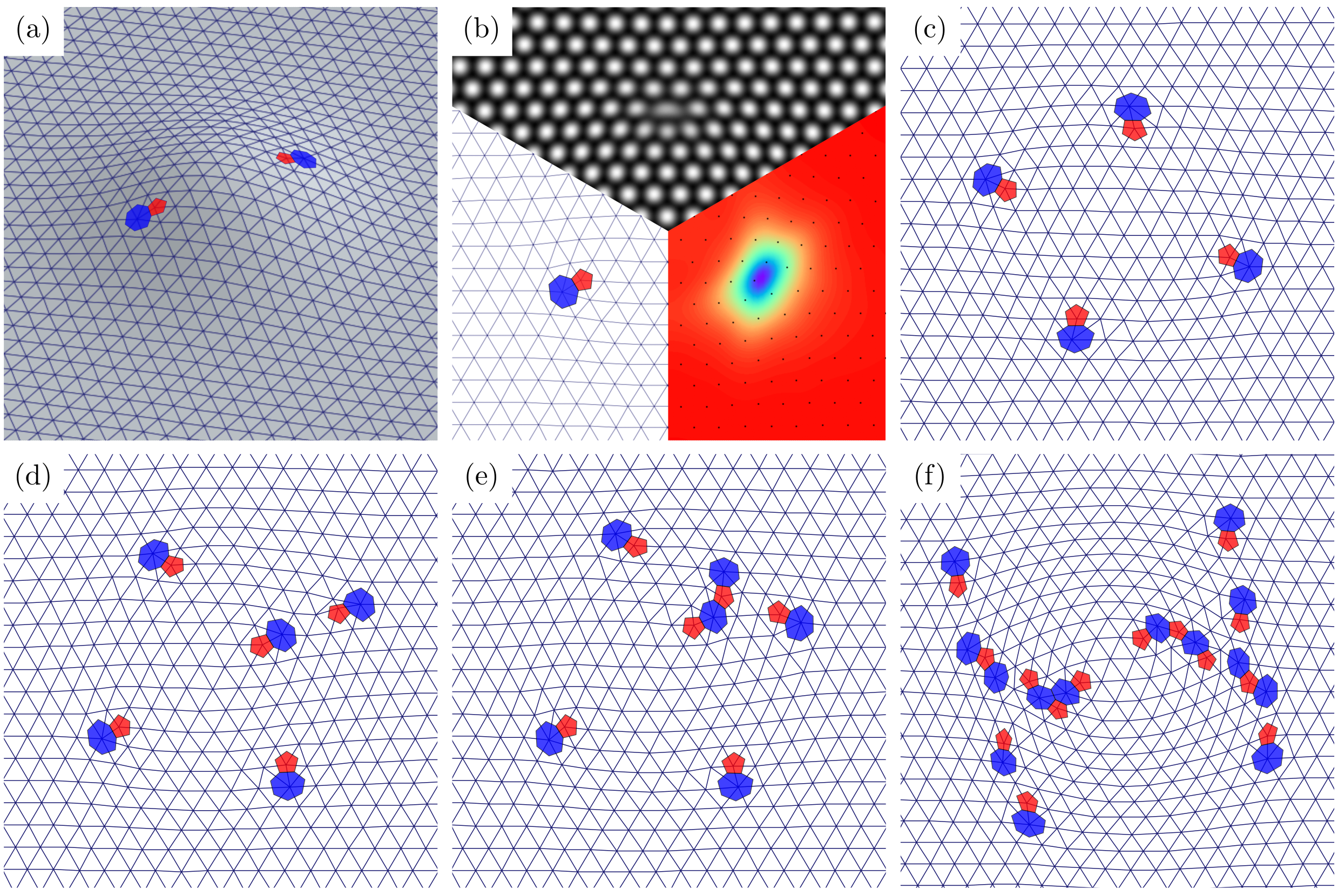}
    \caption{Configurations of defects forming in a crystalline sheet upon its deformation with a Gaussian profile as in Eq.~\eqref{eq:gprof} obtained by the \sAPFC model. The transition from circular defect arrangements (for a number of defects less than or equal to four) to asymmetric arrangements (for more than four defects) is showcased. (a) Delaunay triangulation (DT) of density maxima for a 2-defect configuration in a 3D perspective view; (b) top view of a 3-defect configuration. Clockwise from the top: reconstructed density $\psi$ from the amplitudes \eqref{eq:amp_expansion}, $\Phi=\sum_{n=1}^3|\eta_n|^2$ (blue-to-red color scale), DT of density maxima in top view; (c)-(e) DT of density maxima in top view for 4-, 5-, 6-defect configurations, respectively. (f) DT of a more complex defect network featuring $\sim14$ defects arranged in two symmetric clusters. Aspect ratios $\chi$ of the Gaussian deformation profile for the different panels are (a) 0.7, (b) 0.8, (c) 1.0, (d) 1.0, (e) 1.0, (f) 1.6. In all panels, dislocations are represented by the Voronoi cells of fivefold (red) and sevenfold (blue) disclinations. Energy parameters are set as in Fig.~\ref{fig:bump_apfc}.}
    \label{fig:bump_templates}
\end{figure*}

The deformation energy ($\calF_{\eta,\calS}-\calF_0$ with $\calF_0$ the total energy in a relaxed flat sheet) and the number of defects are plotted as a function of the aspect ratio $\chi$ for the \sAPFC model in Fig.~\ref{fig:bump_apfc}. The globally minimizing configuration is indicated as the continuous line. We also represent local minimizers corresponding to metastable states (reachable by specifying different initial conditions) with dotted lines. The existence of a critical aspect ratio---below which the lattice is frustrated but defect-free and above which the nucleation of defects becomes energetically favorable---is thus made evident. Similarly, the shape of the deformation energy-aspect ratio curve as a smoothly broken power law (\ie a piecewise function given by a sequence of conjoined power laws), further illustrates the competition between elastic and defect energy. Indeed, at a given number of defects, the deformation energy exhibits a minimum, whose value increases with the number of defects. This may be related to the total defect energy, while the U-shaped form of the deformation energies by varying $\chi$ around the minima may be directly connected to the strain energy. As a result, the configuration with fewer defects remains the most stable until it becomes more favorable to pay the energy cost of a defect rather than further deform the lattice.

For a defect-free surface at low aspect ratios ($\chi \lesssim 0.05$), the deformation energy scales as $\chi^4$. This matches an analytical prediction obtained in \cite{Vitelli2006a}, despite our setting not strictly meeting the theoretical requirements. Indeed, the analytical expression was derived in the continuum limit $\rho \ll a$ ($a$ the lattice parameter) and neglecting boundary effects $L \gg \rho$ ($L$ the domain size), while in our simulations, $\rho/a \simeq 4.1$ and $L/\rho \simeq 6.3$. Nevertheless, using Monte Carlo (MC) simulations, the authors have also shown that the analytical results correctly describe the problem beyond their strict validity domain. 
We additionally report the different minimizing defect arrangements in Fig.~\ref{fig:bump_templates}. We note that for four dislocations and less, the minimizing defect configuration consists of a symmetric circular arrangement. This symmetry is lost for five dislocations and more, as the repelling force between dislocations increases. 
These results are in complete agreement with the MC simulations considered in \cite{Hexemer2007}. For very large aspect ratios $\chi >1.5$, we also recover the same branching behavior of dislocations distributing themselves along scars, similarly to what is observed on spherical crystals \citep{Bausch2003,Backofen2011}, thus evidencing the applicability of the proposed model beyond small deformations. 

\begin{figure}[t!]
    \centering
    \includegraphics[width=0.4\textwidth]{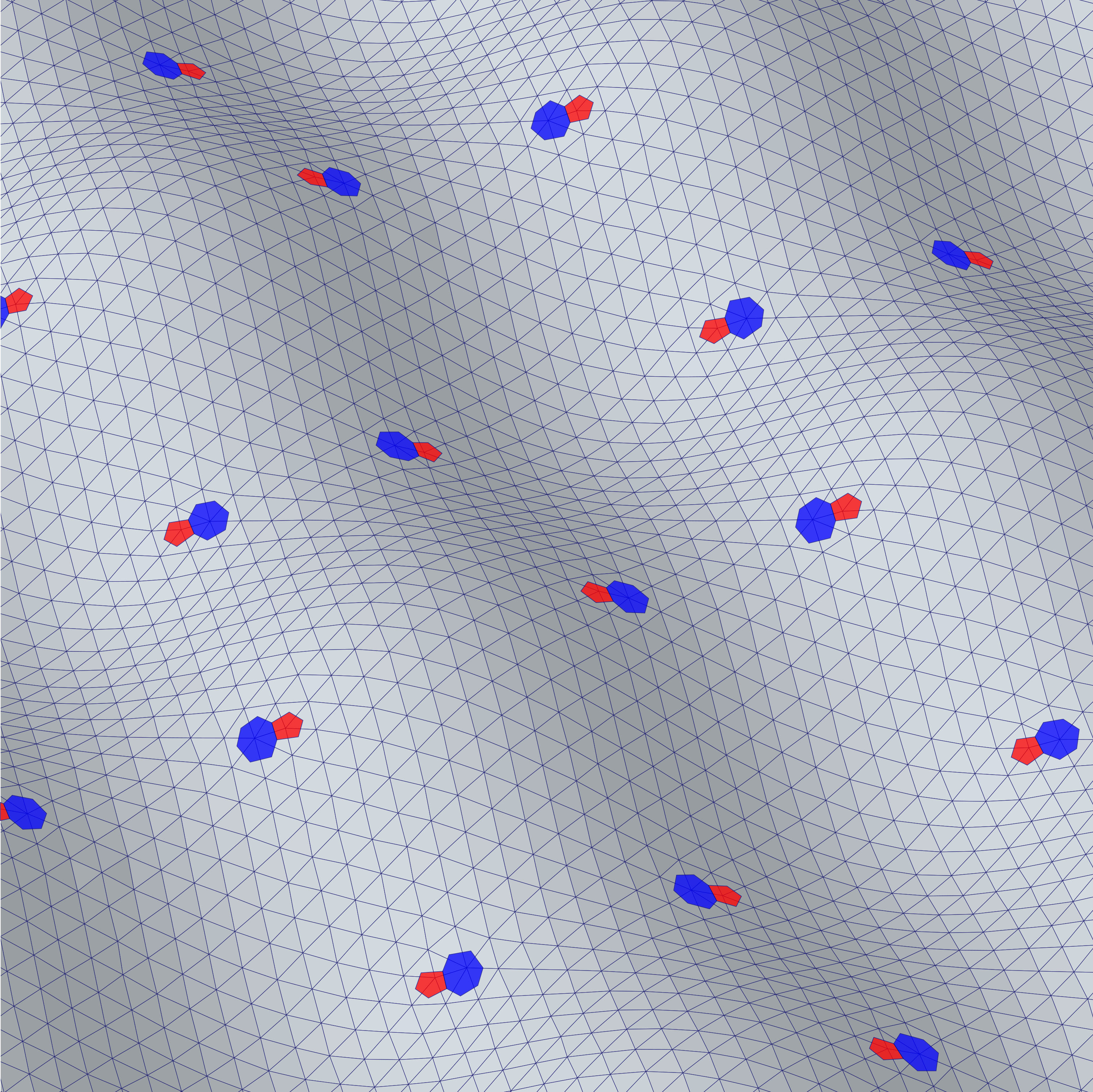}
    \caption{Representative \sAPFC simulation of defect arrangements on a deformed sheet featuring a double sine-wave height profile with aspect ratio $\chi=0.6$. The Delaunay triangulation of density maxima and dislocations are illustrated like in Fig.~\ref{fig:bump_templates}. Energy parameters are set as in Fig.~\ref{fig:bump_apfc}.}
    \label{fig:bump_sine}
\end{figure}

We next consider a double sine-wave surface, \ie
\begin{equation}
    h(x,y)=\chi L_x/(2\pi) \sin(2\pi/L_x x)\sin(2\pi/L_y y).
\end{equation}
featuring a more complex distribution of curvatures and expected behaviors. For low enough aspect ratios, we again obtain defect arrangements consistent with the MC simulations in \cite{Hexemer2007}, see Fig.~\ref{fig:bump_sine}. However, at aspect ratios $\chi > 0.8$, the MC simulations in \cite{Hexemer2007} predict the nucleation of disclinations to reach the global minimum energy state. We remark that these defects cannot be modeled in the (s)APFC frameworks based on the amplitude expansion, Eq.~\eqref{eq:amp_expansion}. Indeed, while these frameworks support defects in the translational symmetry (i.e. dislocations) via phase singularities of the complex amplitudes, they cannot do the same for the rotational symmetry. A different order parameter at the length scale of definition of the amplitudes would be necessary, for instance the one characterizing $p$-atic phases $\phi=|\phi|e^{\i p \theta}$ with $\theta$ a local director field
typically considered for liquid crystals \citep{de1993physics} ($p=6$ for hexatic phases). In contrast, (s)PFC models, which describe microscopic densities, can account for these defects \citep{Backofen2011,Zhang2014b}.

\subsection{Summary of model validation}

The first two subproblems (Sections \ref{sec:buckling} and \ref{sec:surfdislo}) demonstrated the quantitative agreement with established continuous models and previously proposed surface PFC models \citep{Elder2023}. The last subproblem (Section \ref{sec:dislonucleation}) demonstrated that the \sAPFC model also qualitatively captures the essential features of defect nucleation and arrangements on a fixed curved sheet for finite aspect ratios $\chi$. These results go beyond the classical small (ideally infinitesimal) deformation limit. Further, as detailed in \ref{app:validation}, quantitative agreement with the \sPFC model is also excellent in the parameter space of interest, indicating minimal loss of information due to the spatial coarse-graining. These results allow for exploring the full capabilities of the \sAPFC equations, on which we focus for the rest of this paper.

\section{Results}
\label{sec:results}

\subsection{Interplay of surface deformation and dislocations motion}
\label{sec:interaction}

\begin{figure*}
    \centering
    \includegraphics[width=0.95\textwidth]{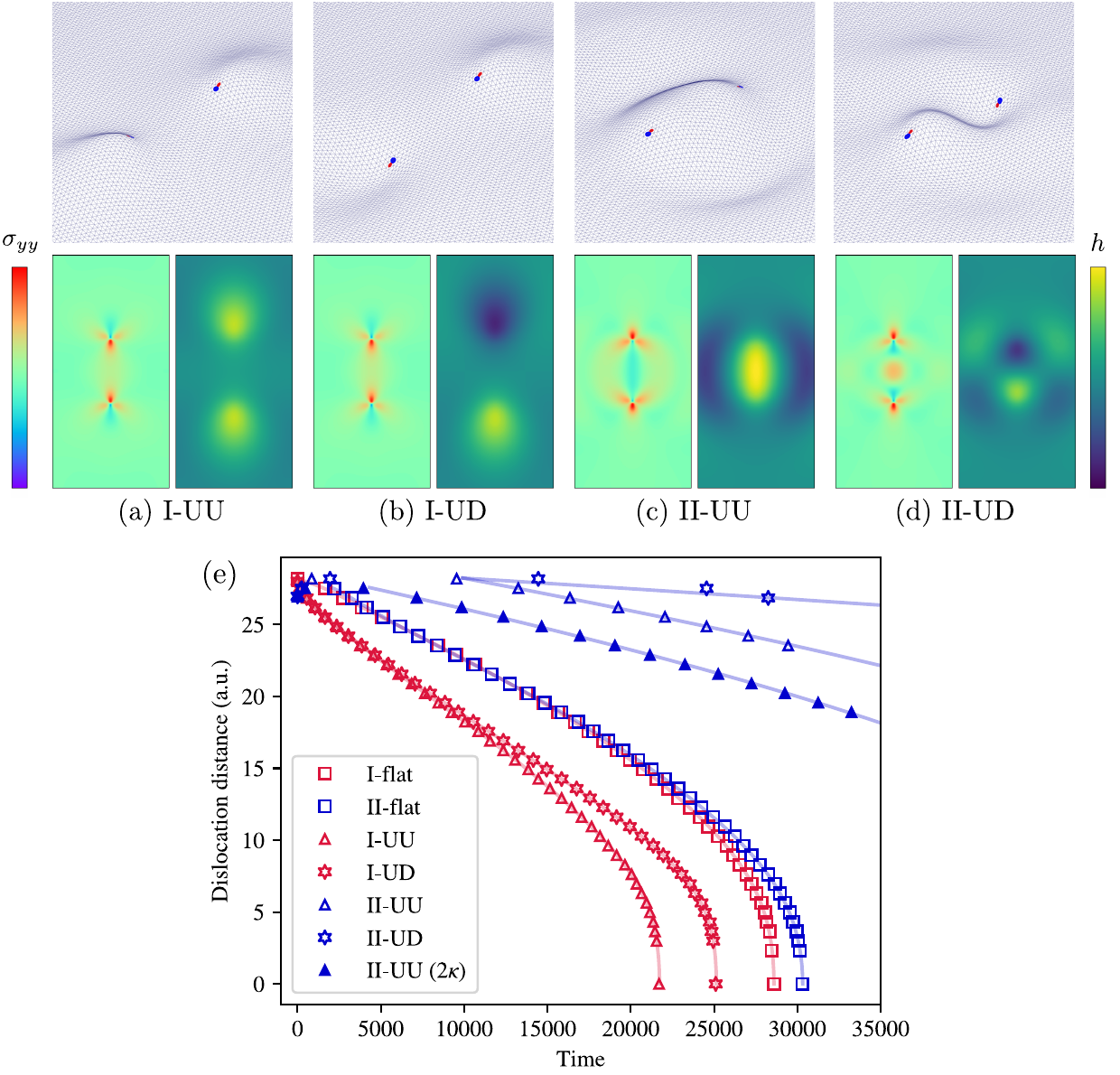}
    \caption{Effects of the out-of-plane deformation on dislocation annihilation (by climbing). (a)-(d) Four representative configurations featuring different combinations of opposite Burgers vector and upward/downward dislocation-induced sheet deformation. Each panel shows: a perspective view of the surface (shown as $4\times h(x,y)$ for illustration purposes) featuring a Delaunay triangulation of density maxima and an illustration of the 5|7 defects (top), $\sigma_{yy}$ (bottom left) and the height profile (bottom right). (e) Dislocation distance as a function of time for the various representative configurations (lines are only guides for the eye). Energy parameters: $A=2, B=0.02, C=-1/2$, and $D=1/3$. Videos of defect motion for I-UU and I-UD illustrating both the Delaunay triangulation of density maxima as in panel (a) (top) and the dislocation density computed as in \cite{Skogvoll2023} are provided as Supplementary Material.}
    \label{fig:disloc_distance}
\end{figure*}

\begin{figure*}
    \centering
    \includegraphics[width=0.98\linewidth]{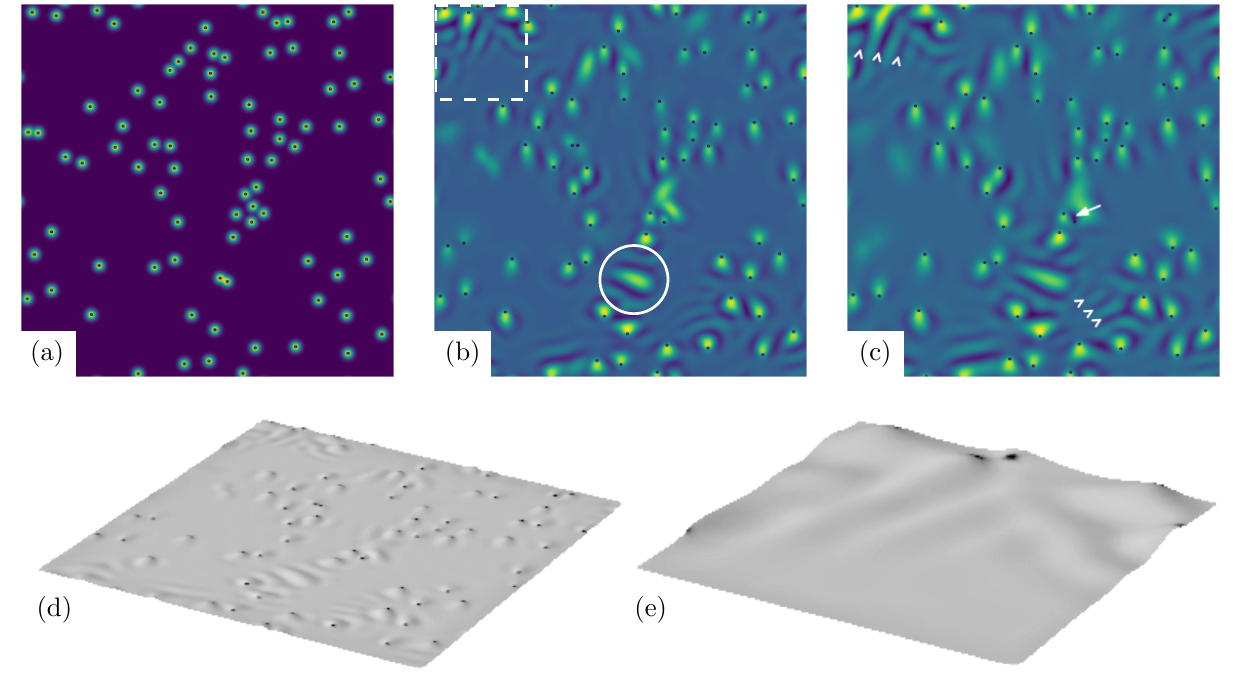}
    \caption{Buckling of a crystalline sheet (side length 340 a.u.) initially hosting 87 dislocations, at times (a) 0, (b) 5000, and (c) 10000. The contour plot shows the height profile. The black dots indicate the dislocations. A few noteworthy features: (\textcircled{}) remnant bulge after dislocation pair annihilation, ($>$) wrinkling, and ($\rightarrow$) upward bulge flipping to a downward bulge. (d)-(e) 3D view of panel (b) and the inset therein. Energy parameters are set as in Fig.~\ref{fig:disloc_distance}.}
    \label{fig:disloc_big}
\end{figure*}

Dislocation interaction is mediated by out-of-plane displacements \citep{C4NR04718D}. To analyze this aspect, we compare via \sAPFC simulations the velocities of a pair of dislocations in different representative configurations of the Burgers vectors and initial surface profiles. 
The amplitude fields are initialized to represent a pair of dislocations 28 atomic units apart in a square domain of linear dimension 171 atomic units via the corresponding displacement field and superimposing the fields from the periodic images \citep{Salvalaglio2020plastic}. 

We look at symmetric configurations where the dislocations are aligned along the $y$ direction. In particular, two configurations are considered: (I) a dislocation at the top (bottom) having Burgers vector parallel (anti-parallel) to the x direction and (II) dislocations as in I but with opposite Burgers vectors. These defects are expected to annihilate by climbing.
Regarding the surface profile, an initial guess consisting of Gaussian bumps centered at each dislocation is chosen. Note that we cannot start from a perfectly flat surface as it represents a stationary condition for the height profile. We consider three cases: i) flat space, ii) upward buckling at each dislocation (denoted by UU), and iii) upward buckling at one dislocation and downward buckling at the other (denoted by UD). These configurations---after relaxation of the initial condition---are illustrated in Fig.~\ref{fig:disloc_distance}(a)--\ref{fig:disloc_distance}(d) via a perspective view of the surface featuring a Delaunay triangulation of density maxima and an illustration of the defects (top), as well as a top view of the surface with color maps representing $\sigma_{yy}$ (bottom left) and the height profile (bottom right). By choosing a high ratio between the height and amplitude mobilities (typically $M_h/M_\eta = 1\times10^5$), we are then able to relax the height and amplitudes to a lower energy configuration while almost completely freezing the movement of the dislocations. Because the system is inherently out of equilibrium, we can not reach a true stationary profile. Instead, this procedure allows us to get a pseudo-stationary height profile, which we use to initialize a simulation where the dislocations can move to study their velocities (with $M_h/M_\eta = 2000$).

The results are shown in Fig.~\ref{fig:disloc_distance}(e). In the flat setting, the dislocations merge towards the center in 28500 and 28000 timesteps in configurations I and II, respectively. Such a slight difference emerges from the asymmetry in the magnitude of compressive/tensile lattice deformations due to nonlinear effects \citep{Huter2016}. Different signs of the stress lobes within the dislocations in the two configurations are obtained, thus driving the motion of the dislocation consistent with the classical Peach-Köhler equation \citep{Skogvoll2022pfc} with slightly different velocities. Out-of-plane deformations contribute to dislocation motion, as anticipated in the previous section. A bulge forms at the region close to the dislocation with negative hydrostatic stress. The dislocation then moves towards the opposite side while dragging the deformed surface. In other, perhaps more visual, words, the dislocations ``surf'' on the bulges they create (see also videos in the Supplementary Material). As a result, when considering the dipoles in a deformable sheet, the dislocations merge faster than in the flat setting when the bulges form opposite to the merging direction, corresponding to the configuration I. On the other hand, when the bulges are between the dislocations, as in configuration II, the movement of the dislocations is considerably slowed down, even moving away from the center initially, as the motion induced by the out-of-plane deformation is opposite to the one dictated by the elastic driving force. 

Focusing on configuration I, we take a closer look at the dynamics of the dislocation merging. At initial times, when the dislocations are far enough from each other, the interactions between their respective bulges are minimal, and the time evolution of the distance is identical in the UU and UD settings. The merging velocity also appears to be greater than in the flat case. Still, we refrain from specific quantitative comments as this result depends on the mobility ratio and the pseudo-stationary surface profile used for initialization. Then, as the dislocations get closer and the bulges they create interact, we see the merging velocity in the UD case decrease relative to its UU counterpart. In short, while an upward bulge drives the dislocation identically to a downward one, as expected from the symmetry of surface deformation for which only the Gaussian curvature is relevant, the interaction of two identical bulges is significantly different from that of opposite bulges. Therefore, besides an additional driving force on a single dislocation, the out-of-plane deformation also mediates an effect on dislocation-dislocation interaction. The simulations for the configurations I-UU and I-UD are also provided as videos in the Supplementary Material.

 Similar conclusions are reached by looking at configuration II. Moreover, the bending stiffness of the deformable sheet is expected to affect this behavior as it directly controls the magnitude of out-of-plane deformation. This is illustrated by considering the II-UU case for two different values of $\kappa$; c.f. II-UU and II-UU (2$\kappa$). Larger values of this parameter lead to less significant out-of-plane deformation and, in turn, a reduced contribution of the bulge-induced motion with the flat surface limit consistently reached for $\kappa \rightarrow \infty$.

We have seen that interactions between dislocations on a deformable sheet are significantly more complex than in a flat (rigid) domain. 
In the following, we additionally show that dislocation interaction in systems hosting many dislocations may also be mediated by surface wrinkling and lead to highly non-trivial behaviors for surface deformations at and close to defects.

\subsection{Initiation of surface wrinkling and other effects resulting from defect interactions}
\label{sec:largescale}

We present a proof of concept of a relatively large-scale simulation enabled by the presented \sAPFC framework. In a square domain of linear dimension 340 a.u., we first introduce (arbitrarily) 256 dislocations at random positions via the appropriate displacement field from linear elastic theory. Without loss of generality for the targeted evidence, and to facilitate the discussion, we only consider Burgers vectors along the $x$-axis. Then, because this initial displacement field from elastic theory is singular, we let the system relax in the flat configuration for a short time to allow the amplitudes to adjust. During this preliminary relaxation, some dislocations annihilate, resulting in a final system containing 87 dislocations. This relaxed configuration is then taken as the initial condition for the fully coupled simulation with out-of-plane displacements---where an upward out-of-plane bump has been introduced at each defect---and we let the system evolve. As dislocations move, driven by both dislocation-induced elastic fields (as in the flat configuration) and out-of-plane deformations, we obtain an overview of all aspects described in previous sections as well as new phenomena, deriving from the complex interactions between dislocations (see Fig.~\ref{fig:disloc_big}). For instance, the various dislocation bulges may interact to create a wrinkled surface. Likewise, while the surface profile at all dislocations was initialized as an upward bulge, we notice that it may flip to a downward bulge due to interaction with another dislocation, which is consistent with the fact that both configurations are indistinguishable in regards to their energies while their interaction may break the underlying symmetry. Additionally, due to the non-trivial dynamics of many dislocations, a combination of the phenomenology presented in Fig.~\ref{fig:disloc_distance} may occur.

\section{Conclusion}
\label{sec:conclusion}
We provided a novel approach to model crystalline sheets at the mesoscale. This has been achieved via a framework for gradient-flow equations---based on a height formulation where surface deformations in the normal direction are considered---that extends the description of crystal lattices conveyed by PFC and APFC models to non-flat surfaces. We focused, in particular, on the latter in several relevant settings. Excellent agreement with atomistic and continuum models was found in both elastic and plastic regimes, demonstrating the ability of the \sAPFC model to accurately describe dislocations on deformable surfaces while solving for slowly varying fields. The realized two-way coupling between the amplitude fields and the surface height profile, where the amplitudes adjust to the surface profile, and long-range elastic fields of defects relax by out-of-plane deformations, allows for describing new phenomena, going beyond the possibilities of previous continuous approaches. 

Considering the velocities of dislocations, we have demonstrated the existence of an additional driving force related to out-of-plane deformation that stirs an isolated dislocation away from the top of the bulge it creates. For interacting dislocations, depending on the nature of the bulges (whether they are pointing in the same direction or not), this additional contribution modifies the classical dynamics leading to pair annihilation in flat domains. This evidence points to a large complexity of behaviors one may expect on crystalline sheets in the presence of many dislocations. Not only do dislocations form bulges on which they "surf", but their interaction also leads to the wrinkling of the crystalline sheet. 

We consciously do not model a specific material but only focus on general coupling mechanisms in generic crystalline sheets. The derived \sAPFC model provides the desired mesoscopic modeling framework, which simultaneously accounts for long-range elastic fields, dislocation motion, localized defect-induced effects, defect-defect interactions, as well as out-of-plane deformations, and responds to local curvature. It thus provides the theoretical basis for comprehensively studying the mechanical properties of crystalline sheets. We remark that applying this concept to specifically model graphene or other 2D materials quantitatively would require adjustments like extended parameterization of the free energy. 
Extensions towards the investigation of crystalline bi- and multi-layers, see e.g. \citep{DaiPRB2016,Elder2023} can be addressed by considering energy terms for each layer and layer-layer interactions. This would enable the study of technologically-relevant systems such as stacked 2D materials like bilayer or multilayer graphene, twisted graphene layers, as well as extensions to heterogeneous multilayers involving for instance hexagonal boron nitride on graphene or Transition Metal Dichalcogenides \citep{guo2021stacking}.

An aspect that would require further, dedicated investigation is the analysis of the height profile after the annihilation of a dislocation pair. Due to the relatively low ratio between the mobilities for the height profile and the amplitudes, arbitrarily chosen to observe dislocations moving and dragging their bulges, the out-of-plane deformations slowly flatten out with time after dislocation annihilation. The corresponding timescale may be controlled with the mobility ratio and varied to explore different regimes. In addition, we consider here a purely diffusive dynamics for the amplitudes that cannot capture fast elastic response. Formulations accounting for this regime have been proposed for the APFC model in flat space \citep{Heinonen2016,Salvalaglio2020plastic}. However, the extension of these approaches to deformable sheets in the formulation considered here is non-trivial. It will be the object of future work together with quantitative inspections of dislocation interactions.



\section*{Data availability}

The data that support the findings of this study are openly available at the following URL/DOI: \href{https://doi.org/10.5281/zenodo.13150877}{10.5281/zenodo.13150877}

\section*{Acknowledgments}
M.S. acknowledges support from the Emmy Noether Program of the German Research Foundation (DFG) - project number 447241406. M.S. and A.V.  acknowledge support from the German Research Foundation (DFG) within FOR 3013 - project number 417223351. The authors gratefully acknowledge the Center for Information Services and High-Performance Computing [Zentrum für Informationsdienste und Hochleistungsrechnen (ZIH)] at TU Dresden for computing time.

\appendix
\setcounter{figure}{0}

\section{Height formulation (Monge patches)}
\label{app:mongepatches}
Consider the domain $ \Omega \subset \mathbb{R}^2 $, a sufficiently smooth height function $h: \Omega \rightarrow \mathbb{R}$ and parametrization $\pmb{X}:\Omega \rightarrow \calS \subset \mathbb{R}^3$ of the form: $\pmb{X}(x, y) = [x, y, h(x, y)]$.
We call surfaces embeddable by such a parametrization \textit{Monge patches}. 
We denote by the usual symbol $\nablaf$ the flat derivative, \ie the gradient in $\mathbb{R}^2$, such that $\nablaf f = (\partial_x f)\pmb{e}_x + (\partial_y f)\pmb{e}_y$.
with $f$ an arbitrary scalar field and $\{\mathbf{e}_i\}$ the Cartesian basis. 
The metric tensor, \ie the first fundamental form, then reads:
\begin{equation}
    \pmb{g} = 1 + (\nablaf h)^2, \qquad \bigg(\ g_{ij}=\delta_{ij}+(\partial_ih)(\partial_jh)\ \bigg).
    \label{eq:metric}
\end{equation}
We also define $\g=1+(\partial_x h)^2+(\partial_y h)^2$ and $\langle \ \cdot \ ,\ \cdot \ \rangle$ as the inner product in $\Omega$. With these notions, we can specify quantities on the surface, denoted with the subscript $\calS$, in terms of their definition in $\Omega$ and $h$. Useful quantities used in this work are
the norm 
\begin{equation}\label{eq:appNorm}
\begin{split}
    ||\mathbf{w}||_\calS&=||\mathbf{w}||-\frac{\langle \nabla h, \mathbf{w}\rangle^2}{\g}, \\ \bigg(\ &=  w_x^2+w_y^2-\frac{(w_x\partial_x h+w_y\partial_y h)^2}{\g}\ \bigg),
    \end{split}
\end{equation}
the surface normal vector
\begin{equation}\label{eq:appSurfaceNormal}
    \pmb{\nu}=
      \frac{1}{\sqrt{|\pmb{g}|}}
    \begin{bmatrix}
           -\dx h \\
           -\dy h \\
           1
         \end{bmatrix}_{\mathbb{R}^3},
\end{equation}
the mean curvature
\begin{equation}\label{eq:appH1}
\begin{split}
&\calH = \frac{1}{\sqrt{\g}}\left(\lap h - \frac{\scal{(\nablaf h)^2}{\nablaf\nablaf h}}{\g}\right), \qquad \qquad \\
&\left(\ =\frac{\partial_{xx} h+\partial_{yy} h}{\sqrt{\g}}-\frac{(\partial_{xx} h)(\partial_x h)^2+(\partial_{yy} h)(\partial_y h)^2}{\g^{3/2}}\ \right.\\ & \left. -\frac{2(\partial_{x y} h)(\partial_x h)(\partial_y h)}{\g^{3/2}}\right),
\end{split}
\end{equation}\label{eq:appGradS1}
the Gaussian curvature
\begin{equation}\label{eq:appH2}
\mathcal{K} = \frac{|\nablaf^2 h|}{\g^2}, \qquad 
\left(\ =\frac{(\partial_{xx} h)(\partial_{yy} h)-(\partial_{xy}h)^2}{\g^2} \right),
\end{equation}\label{eq:appGradS2}
the surface gradient
\begin{equation}
\begin{split}
    &\nabla_\calS f = \pmb{g}^{-1} \cdot \nabla f = \nabla f - \frac{\langle \nabla h,\nabla f\rangle}{|\pmb{g}|} \nabla h, \\
    &\left(\ f^{|i} = \partial_i f-\frac{(\partial_x f)(\partial_x h)+(\partial_y f)(\partial_y h)}{|\pmb{g}|}\partial_i h\ \right),
\end{split}
\end{equation}
the surface divergence
\begin{equation}\label{eq:appDivS}
\begin{split}
{\rm div}_\calS \mathbf{w} =& \nablaf \cdot \mathbf{w} + \frac{\scal{(\nablaf\nablaf h)\nablaf h}{\mathbf{w}}}{\g}, \\
& \left(\ =\partial_x w^x + \partial_y w^y + \frac{w^x(\dx h)\dx^2 h + w^y(\dy h)\dy^2 h
}{\g} \ 
\right.\\ & \left. +
\frac{ (w^x \dy h + w^y \dx h)\partial_{xy} h}{\g}
\right),
\end{split}
\end{equation}
and the surface Laplacian, or Laplace-Beltrami operator
\begin{equation}\label{eq:appLapS}
\begin{split}
\lapbel  f =& \lap f - \frac{\scal{\nablaf h}{\nablaf f}}{\g}\lap h - \frac{\scal{(\nablaf h)^2}{\nablaf\nablaf f}}{\g} \\&+ \frac{\scal{(\nablaf h)^2}{\nablaf\nablaf h}\scal{\nablaf h}{\nablaf f}}{\g^2},
\\
&\bigg(\ =\partial_{xx}f+\partial_{yy}f
-
\frac{(\partial_xf)(\partial_xh)+(\partial_y f)(\partial_yh)}{\g}(\partial^2_x h+\partial^2_yh)
\\ &
-\frac{(\partial_{xx}f)(\partial_xh)^2+(\partial_{yy}f)(\partial_yh)^2+2(\partial_{xy}f)(\partial_xh)(\partial_yh)}{\g} 
\\
&  +\frac{[(\partial_xf)(\partial_xh)+(\partial_y f)(\partial_yh)][(\partial_{xx}h)(\partial_xh)^2+(\partial_{yy}h)(\partial_yh)^2]}{\g^2}
\\
&  +\frac{[(\partial_xf)(\partial_xh)+(\partial_y f)(\partial_yh)][2(\partial_{xy}h)(\partial_xh)(\partial_yh)]}{\g^2}\ \bigg).
\end{split}
\end{equation}
Denoting $\pmb{II}$ the shape operator, $\Pi_{QS}\pmb{II}$ its trace-free part, and $\Gamma_{ij}^{k}$ the Christoffel symbol of the second kind, we collect below a few useful identities:
\begin{equation}
II_{ij} = \frac{\partial_{ij} h}{\sqrt{\g}}, 
\end{equation}
\begin{equation}
\Pi_{QS}\pmb{II} = \pmb{II}-\frac{\calH}{2}\mathrm{Id}_{\rmT\calS},
\end{equation}
\begin{equation}
\Gamma_{ij}^{k} = \frac{1}{\g}(\partial_{ij}h)\partial_k h,
\end{equation}
\begin{equation}
\dS(\pmb{g}^{-1}) = 2\pmb{II}\delta\xi,
\label{eq:ds_ginv}
\end{equation}
\begin{equation}
\dS(\Gamma_{ij}^k) = [\nablas(\delta\xi)\otimes\pmb{II} - \pmb{II}\otimes\nablas(\delta\xi)]_{ij}^k - [\nablas(\delta\xi\,\pmb{II})]_{ij}^k,
\label{eq:ds_christoffel}
\end{equation}
\begin{equation}
\dS\left(\int_\calS f\rmd\calS\right) = \int_\calS (\dS f - \calH f\delta\xi)\rmd\calS,
\label{eq:ds_int}
\end{equation}
\begin{equation}
\dS\calH = \lapbel(\delta\xi) + (\calH^2-2\calK)\delta\xi
\label{eq:ds_meancurv},
\end{equation}
\begin{equation}
\begin{split}
\dS (\lapbel f) =& \lapbel(\dS f) + 2\scal{(\Pi_{QS}\pmb{II})\nablas f}{\nablas(\delta\xi)}_\calS \\ &+ \left(\scal{\nablas f}{\nablas\calH}_\calS+2\scal{\lapbel f}{\pmb{II}}_\calS\right)\delta\xi,
\label{eq:ds_lapbel}
\end{split}
\end{equation}
\begin{equation}
\dS||\nablas f||_\calS^2 = 2\left(\scal{\nablas f}{\nablas(\delta_\calS f)}_\calS + \scal{(\nablas f)^2}{\pmb{II}}_\calS\delta\xi\right),
\end{equation}
with $\langle \  \cdot \ , \ \cdot \ \rangle_\mathcal{S}$ the inner product on $\calS$ and $f$ a generic function defined on $\Omega$. For further details on how these expressions are obtained, we refer to \cite{Nitschke2020}.

\section{Functional derivatives}
\label{app:FuncDer}

\subsection{Bending energy}
\label{app:Bending}
The variation of the bending energy, which depends on the curvature and the surface parametrization, is the same in all the models considered in this work. By combining \eqref{eq:ds_int} and \eqref{eq:ds_meancurv} we obtain the following expression
\begin{equation}
    \dS\calF_b = \kappa\int_\calS\calH\left[\lapbel(\delta\xi) + \left(\frac{\calH^2}{2}-2\calK\right)\delta\xi\right]\rmd\calS.
\end{equation}
Then, with convenient boundary conditions, \ie such that all line integrals $\int_{\partial\calS}$ are zero,
\begin{equation}
    \frac{\delta\calF_b}{\delta\xi} = \kappa\left[\lapbel\calH + \calH\left(\frac{\calH^2}{2}-2\calK\right)\right].
    \label{eq:dFb_dxi}
\end{equation}

\subsection{sPFC energy}
\label{app:PFC}

In the \sPFC model, 
we need to compute variations of $\mathcal{F}_{\psi,\calS}$. 
Under $\dS\psi=0$, implying that the local values of the density field are not affected by surface deformations a priori (while gradients are), the variation of the elastic energy term $F_{\calL}=\int_\calS\frac{A}{2}\psi(\calLs\psi)\,\rmd\calS=\int_\calS\frac{A}{2}(\calM_\calS\psi)^2\,\rmd\calS$ reads:
\begin{equation}
    \dS\calF_{\calL} = \int_\calS \left[A(\calM_\calS\psi)\dS(\lapbel\psi) - \calH\left(\frac{A}{2}\psi\calLs\psi \right)\delta\xi\right]\rmd\calS,
\end{equation}
where $\calM_\calS=q^2+\lapbel$. The functional derivative with respect to normal variations of the surface profile is then
\begin{equation}
\begin{split}
    \frac{\delta\calF_{\cal{L}}}{\delta\xi} =  &-2A\,\mathrm{div}_\calS \left[(\calM_\calS\psi)(\Pi_{QS}\pmb{II}\nablas\psi) \right] - \calH\left(\frac{A}{2}\psi\calLs\psi\right)
    \\
    &+ A(\calM_\calS\psi)\left(\scal{\nablas\psi}{\nablas\calH}_\calS+2\scal{\lapbel\psi}{\pmb{II}}_\calS\right). 
    \label{eq:pfc_dF_dxi}
\end{split}
\end{equation}
The variation of $\mathcal{F}_{\psi,\calS}$ with respect to $\psi$ results analogous to the classical PFC model, with adapted differential operators accounting for proper derivatives on the surface:
\begin{equation}
    \frac{\delta\calF_{\psi,\calS}}{\delta\psi} = A\calLs\psi + B\psi + C\psi^2 + D\psi^3.
    \label{eq:pfc_dF_dpsi}
\end{equation}

\subsection{sAPFC energy}
\label{app:APFC}

In the \sAPFC model, we consider variations of $\mathcal{F}_{\eta,\calS}$ and recall that $\eta_n$ are complex functions. We begin by looking at the variation of the energy term, including the differential operator, namely the elastic energy term $\calF_\calG=A\sum_{n=-N}^N\int_\calS|\calG_n^\calS\eta_n|^2\rmd\calS$. Using Eqs.~\eqref{eq:ds_ginv}, \eqref{eq:ds_christoffel}, and \eqref{eq:ds_lapbel}, we have 
\begin{equation}
\begin{split}
    \dS(\calG_n^\calS\eta_n) =& 2\scal{(\Pi_{QS}\pmb{II})(\nablas\eta_n + \i\eta_n\kv_n)}{\nablas(\delta\xi)}_\calS \\ &+2\scal{\pmb{II}}{\nablas(\nablas\eta_n+\i\eta_n\kv_n)}_\calS\delta\xi \\ &+ \scal{\nablas\calH+2\i\pmb{II}\kv_n}{\nablas\eta_n+\i\eta_n\kv_n}_\calS\delta\xi,
\end{split}
\end{equation}
Introducing the operation $(a,b):=\mathrm{Re}(a^* b)$ for all $a,b \in \mathbb{C}$, so that
\begin{equation}
\begin{split}
    \dS\left|\calG_n^\calS\eta_n\right|^2 &= (\calG_n^\calS\eta_n)\,\dS(\calG_n^\calS\eta_n)^* + (\calG_n^\calS\eta_n)^*\,\dS(\calG_n^\calS\eta_n) \\ &= 2\,\mathrm{Re}\boldsymbol{\Big(}(\calG_n^\calS\eta_n)^*\,\dS(\calG_n^\calS\eta_n)\boldsymbol{\Big)}=2\boldsymbol{\Big(}\calG_n^\calS\eta_n,\dS(\calG_n^\calS\eta_n)\boldsymbol{\Big)},
\end{split}
\end{equation}
we may finally compute

\begin{equation}
\begin{split}
    \dS\calF_\calG = A\sum_{n=-N}^N\int_\calS \Big[
      &4\scal{\Big(\calG_n^\calS\eta_n,(\Pi_{QS}\pmb{II})(\nablas\eta_n+\i\eta_n\kv_n)\Big)}{\nablas(\delta\xi)}_\calS \\
    + &4\Big(\calG_n^\calS\eta_n,\scal{\pmb{II}}{\nablas(\nablas\eta_n+\i\eta_n\kv_n)}_\calS\Big)\delta\xi \\
    + &2\Big(\calG_n^\calS\eta_n,\scal{\nablas\calH+2\i\pmb{II}\kv_n}{\nablas\eta_n+\i\eta_n\kv_n}\Big)\delta\xi \\
    - &\calH|\calG_n^\calS\eta_n|^2\delta\xi
    \Big]\rmd\calS,
\end{split}
\end{equation}
so that, similarly to the previous sections, with appropriate boundary conditions (all line integrals taken to be zero), we get
\begin{equation}
\begin{split}
    \frac{\delta\calF_\calG}{\delta\xi} = A\sum_{n=-N}^N\Bigg[\vphantom{\left|\calG_n^\calS\eta_n\right|^2}
    - &4\mathrm{div}_\calS\left(\calG_n^\calS\eta_n,(\Pi_{QS}\pmb{II})(\nablas\eta_n+\i\eta_n\kv_n)\right) \\
    + &4\left(\calG_n^\calS\eta_n,\scal{\pmb{II}}{\nablas(\nablas\eta_n+\i\eta_n\kv_n)}_\calS\right) \\
    + &2\left(\calG_n^\calS\eta_n,\scal{\nablas\calH+2\i\pmb{II}\kv_n}{\nablas\eta_n+\i\eta_n\kv_n}\right)
    \vphantom{\left|\calG_n^\calS\eta_n\right|^2}
    \\ - &\calH\left|\calG_n^\calS\eta_n\right|^2 \Bigg],
    \label{eq:dFg_dxi}
\end{split}
\end{equation}
%
\ie explicitly in the height formulation:
\begin{equation}
\begin{split}
    \frac{\delta\calF_\calG}{\delta\xi}& = A\sum_{n=-N}^N\Bigg[\vphantom{\left|\calG_n^\calS\eta_n\right|^2}
    - 4\mathrm{div}_\calS\left(\calG_n^\calS\eta_n,\overline{\overline{\Pi_{QS}\pmb{II}}}(\nablaf\eta_n+\i\eta_n\kv_n)\right) \\
    + &4\left(\calG_n^\calS\eta_n,\scal{\overline{\overline{\pmb{II}}}}{\nablaf^2\eta_n+\i\kv_n\otimes\nablaf\eta_n-\frac{\scal{\nablaf h}{\nablaf\eta_n+\i\eta_n\kv_n}}{\g}\nablaf^2 h}\right) \\
    + &2\left(\calG_n^\calS\eta_n,\scal{\left(\pmb{I}-\frac{(\nablaf h)^2}{\g}\right)\nablaf\calH + 2\i\overline{\overline{\pmb{II}}}\kv_n}{\nablaf\eta_n+\i\eta_n\kv_n}\right)
    \vphantom{\left|\calG_n^\calS\eta_n\right|^2} \\ 
    - &\calH\left|\calG_n^\calS\eta_n\right|^2  \Bigg],
    \label{eq:dFg_dh}
\end{split}
\end{equation}
where
\begin{equation}
\begin{split}
    \calG_n^\calS\eta_n = &\phantom{+}\lap\eta_n - \frac{\scal{\nablaf h}{\nablaf \eta_n}}{\g}\lap h - \frac{\scal{(\nablaf h)^2}{\nablaf\nablaf\eta_n}}{\g} \\& + \frac{\scal{(\nablaf h)^2}{\nablaf\nablaf h}\scal{\nablaf h}{\nablaf\eta_n}}{\g^2}
    \\& +\left(q^2-||\kv_n||^2 + \frac{\scal{\kv_n}{\nablaf h}^2}{\g}\right)\eta_n \\& + 2\i\left(\scal{\kv_n}{\nablaf\eta_n} - \frac{\scal{\nablaf h}{\kv_n}\scal{\nablaf h}{\nablaf\eta_n}}{\g}\right)
    \\
    &-\i\frac{\eta_n}{\g}\left(\scal{\nablaf h}{\kv_n}\lap h - \frac{\scal{(\nablaf h)^2}{\nablaf\nablaf h}\scal{\nablaf h}{\kv_n}}{\g}\right),
    \\
    \overline{\overline{\Pi_{QS}\pmb{II}}} = &\phantom{+}\frac{\pmb{I}-\frac{(\nablaf h)^2}{\g}}{\sqrt{\g}}\left(\pmb{I}-\frac{(\nablaf h)^2}{\g}\right)\lap h \\& - \frac{\pmb{I}-\frac{(\nablaf h)^2}{2\g}}{\sqrt{\g}}\Bigg(\lap h - \frac{\scal{\nablaf\nablaf h}{(\nablaf h)^2}}{\g}\pmb{I}\Bigg),
    \\
    \overline{\overline{\pmb{II}}} = &\phantom{+}\left(\pmb{I}-\frac{(\nablaf h)^2}{\g}\right)\frac{\nablaf\nablaf h}{\sqrt{\g}}\left(\pmb{I}-\frac{(\nablaf h)^2}{\g}\right).
\end{split}
\end{equation}
$\delta\calF_{\eta,\calS}/\delta\eta_n^*$ and $\delta\calF_{\eta,\calS}/\delta\no$ follow the standard derivation:
\begin{equation}
\begin{split}
    \frac{\delta\calF_{\eta,\calS}}{\delta\eta_n^*} =& A\left(\calG_n^\calS\right)^2\!\!\eta_n + \frac{\partial g^s}{\partial\eta_n^*}, \\
    \frac{\delta\calF_{\eta,\calS}}{\delta\no} =& (C+3D\no)\zeta_2 + D\zeta_3 + A\no + C\no^2 + D\no^3.
   \end{split}
\end{equation}

\section{Numerical methods}
\label{app:numerics}
To solve the sPFC and sAPFC models we consider a Fourier pseudo-spectral method \citep{Salvalaglio2022ov}. In short, the equations to solve for $\eta_j(\mathbf{r},t)$,  $h(\mathbf{r},t)$, $\psi(\mathbf{r},t)$ or $\overline{\psi}(\mathbf{r},t)$---generically written as
\begin{equation}
\partial_t{f} = {\cal L} f + {\cal N}(f),
\end{equation}
with $f$ the unknown, ${\cal L}$ the linear part in $f$ of the operator, and ${\cal N}$ the nonlinear part---are solved in Fourier space. In other words, we solve for 
\begin{equation}
\partial_t{\widehat{f}_k} = {\cal L}_k \widehat{f}_k + \widehat{{\cal N}}_k,
\label{eq:ode}
\end{equation}
with $\widehat{f}_k$ the Fourier transform of $f$, $\widehat{\cal N}_k$ the Fourier transform of ${\cal N}(f)$ and ${\cal L}_k$ the Fourier transform of ${\cal L}$ resulting in an algebraic expression of the wave vector. The solution at $t+\Delta t$, with $\Delta t$ the timestep, is then obtained via an inverse Fourier transform of $\widehat{f}_k(t+\Delta t)$ computed by an exponential integrator step \citep{Certaine1960,Pope1963}.
\begin{equation}
\widehat{f}_{k}(t+\Delta t) 
\approx\ {\rm e}^{{\cal L}_k\Delta t}\widehat{f}_k(t)
+\frac{{\rm e}^{{\cal L}_k\Delta t}-1}{{\cal L}_k} \widehat{N}_k(t).
\label{eq:spec}
\end{equation}
We recall that with this method, periodic boundary conditions are considered.
An established Fast-Fourier Transform algorithm (FFTW) is exploited \citep{Frigo2005}. Using this simple approach is enough for the goals of this work and, at the same time, showcases the convenience of the reported model to account for surface deformations. Notice, however, that there are no restrictions in considering more efficient approaches leveraging peculiar features of the models, e.g., adaptive frameworks proposed for efficient simulations of APFC models in real space \citep{Praetorius2019}.

\section{Comparison of sPFC and sAPFC models}
\label{app:validation}
\subsection{Buckling under a compressive load}
\label{app:validation-buckling}

\begin{figure*}[]
    \centering
     \includegraphics[width=0.98\textwidth]{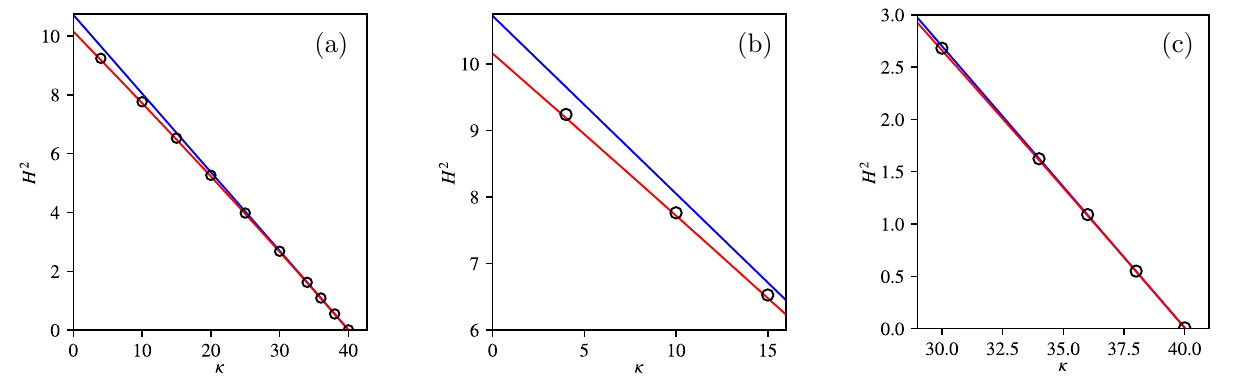}
    \caption{Assessment of the buckling behavior encoded in the \sPFC model. (a) Simulation results (black circles) from \cite{Elder2021} in terms of $H^2$ (see parametrization in Fig.~\ref{fig:buckling_schematic}) against bending stiffness $\kappa$ for $\varepsilon=-0.0219$, compared with the highest-order analytical expression of the energy with (red, Eq.~\eqref{eq:correction}) and without (blue, Eq.~\eqref{eq:highorder}) the approximation discussed in~\ref{app:validation-buckling}. (b) and (c) show details of the plot in (a) at low and large bending stiffness, respectively.}
    \label{fig:buckling_pfc}
\end{figure*}

Figure~\ref{fig:buckling_analytic} in the main text illustrated \sAPFC simulation results of buckling in terms of $H^2$ (with $H$ the maximum height, see Fig.~\ref{fig:buckling_schematic}) by varying the bending stiffness $\kappa$. We also compared these results to various analytical approximations. 
The highest-order analytical prediction, which accurately reproduces \sAPFC results as discussed in the main text, is also expected to describe well buckling obtained by the \sPFC model. Indeed, to describe elasticity in the PFC model, the standard approach consists of considering the amplitude formulation \cite{Salvalaglio2022ov} and, thus, the \sAPFC model in the first place. However, in previous formulations coupling the PFC to out-of-plane deformation, the differential operator and the bending energy are approximated by their leading order in terms of the surface height gradients. As shown in Fig.~\ref{fig:buckling_pfc}, a clear difference is visible in the region of large buckling height between the results of numerical simulations considering such approximations from \cite{Elder2021} (black circles) and the analytic expressions derived from the \sAPFC model (blue solid line). By modifying the analytical derivations to account for the simplifications considered in \cite{Elder2021}, the predicted buckling height relation reads:
\begin{align}
\begin{split}
    (HQ)^2 = &-4(\eps-\eps_0) - \frac{19}{2}\left(\eps^2+\eps_0^2-\frac{58}{19}\eps_0\eps\right) \\
    &- \frac{361}{8}\eps^3 + \frac{1177}{6}\eps^2\eps_0 - \frac{4235}{24}\eps_0^2\eps \\
    &+ \frac{509}{12}\eps_0^3\left(1-\frac{32}{152}\frac{\phi_1 + 8\phi_2}{\phi}\right)+ \bigO(\eps_0^4),
\end{split}
\end{align}
and we recover an excellent agreement with the corresponding data of \cite{Elder2021} (see black circles and red solid line in \ref{fig:buckling_pfc}). Interestingly, looking back at Eq.~\eqref{eq:hq} as a function of $\kappa$:
\begin{equation}\label{eq:correction}
    (HQ)^2 = -4\eps - \frac{4\kappa Q^2}{9A\phi^2}\left(1+\frac{5}{2}\eps\right) 
    \simeq -4\eps - \frac{4\kappa Q^2}{9A\left[\left(1 - \frac{5}{4}\eps\right)\phi\right]^2},
\end{equation}
we observe that the higher-order $\bigO(\eps_0^3,g^3)$ expression matches the $\bigO(\eps_0^2,g^3)$ expression if the amplitude is corrected by a factor $1 - (5/4)\eps$, see the denominator of the final expression in Eq.~\eqref{eq:correction}. For $\eps=-0.0219$, this quantity $1-(5/4)\eps=1.027$ matches the correcting factor of $1.031$ proposed in \cite{Elder2021} to have the lower-order approximation fit the simulation results.

With the analysis of the buckling behavior from Section~\ref{sec:buckling} and above, we may conclude that: i) the \sAPFC model matches well with known results concerning PFC modeling of deformable surface in elastic regimes, and ii) it encodes a description of elasticity beyond infinitesimal deformations. Moreover, we reported here a rigorous analysis shedding new light on the correction proposed in \cite{Elder2021}, which actually serves to compensate for the low-order expansion in the critical strain.
Additionally, we remark on two important aspects of considering the \sAPFC model instead of the \sPFC model. Without any defect to resolve, the \sAPFC has a significant computational advantage as a very coarse mesh, which only needs to resolve the slowly varying height profile, can be used. For instance, the \sAPFC simulations are conducted with $\delta x = \delta y \simeq 31.4$ (as well as $15.7$ and $7.9$ to verify convergence). In contrast, the \sPFC simulations required $\delta x = \delta y \simeq 0.48$ to correctly resolve the microscopic density field $\psi$. Moreover, amplitudes can be directly used to compute the elastic field \citep{Salvalaglio2019,Salvalaglio2020plastic} without using additional filtering and numerical coarse-graining procedures \citep{Skogvoll2021,Elder2021}.

\subsection{Defect arrangements on a fixed surface profile}

\begin{figure}[t]
    \centering
    \includegraphics[width=0.85\linewidth]{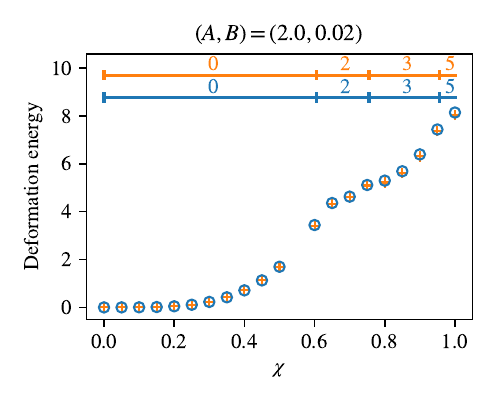}
    \includegraphics[width=0.85\linewidth]{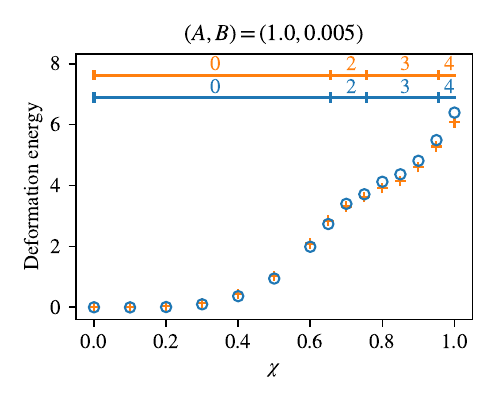}
    \caption{Deformation energy ($\calF_{\eta,\calS}-\calF_0$) for the system in Fig.~\ref{fig:bump_templates} as a function of aspect ratio for both the \sPFC model (blue circles) and \sAPFC model (orange crosses). The two plots show the results obtained for $(C, D)=(-1/2,1/3)$ as in Fig.~\ref{fig:bump_apfc}, with two choices of the energy parameters $A$ and $B$ reported in the panels (see definition of the energy in Eq.~\eqref{eq:F_APFC}). The number of defects in the energy-minimizing configuration for different $\chi$, coinciding for the two considered models, is also reported; see the segmented lines on the top part of the plots (colors correspond to the simulation results).}
    \label{fig:bump_pfc_apfc}
\end{figure}

Since the \sAPFC model is obtained as a coarse-grained \sPFC model, we now present comparative results of the two models in Fig.~\ref{fig:bump_pfc_apfc} to inspect the validity of the coarse-graining approximation more specifically and in the presence of defects. The agreement between these models is almost perfect for both the energy and the number of dislocations. This further confirms the validity of the coarse-graining underlying the devised mesoscale framework. In particular, using the flat scalar product approximation on curved substrates (see also discussion in Section~\ref{sec:sAPFC}) does not introduce spurious effects in the presence of both lattice deformation and dislocation nucleation, \ie the phenomena we aim to describe with our approach. We note that this holds in a setting where the maximum height gradients reach $0.61$ for $\chi=1$, which is far beyond the small deformation limit. Nevertheless, for $\chi>0.5$, the surface energy is observed to be slightly higher ($1\%$ at most) in the \sPFC simulations compared to the \sAPFC results, which might be ascribed to that approximation. We find that the number of defects and the energy depends on the energy parameters $A$, namely controlling elastic constants, and $B$ corresponding to the quenching depth \citep{Skogvoll2023} (the latter being the parameter controlling the order-disorder/solid-liquid phase transition for $B\lessgtr B_c$, with $B_c$ the critical point). In addition, the agreement illustrated in Fig.~\ref{fig:bump_pfc_apfc} holds for relatively large values of $A$ (typically $A \ge 2$). At lower values of $A$, the cohesion between atoms effectively described by the PFC model becomes very weak, mimicking a colloidal behavior, which is naturally not reproducible by APFC models, as it enforces a lattice structure through Eq.~\eqref{eq:n_app}. Nevertheless, since our focus is on modeling crystalline structures, the comparison at these low $A$ values is not meaningful.

When sufficiently close to the solid-liquid transition ($B\sim B_c$), we observe that partial melting at the top of the bump is the most energetically favorable state with the \sPFC model, whereas the \sAPFC with the approximation of constant average density $\overline{\psi}$ still predicts a fully crystallized phase. This results from neglecting Eq.~\eqref{eq:barpsi}, as further discussed below. Aiming here at describing thin sheets in the crystalline state, we however propose to use the \sAPFC model with constant $\overline{\psi}$ as it provides an accurate coarse-graining as seen in Fig.~\ref{fig:bump_pfc_apfc}. An explicit comparison between the solution of the full system of equations \eqref{eq:hapfc}--\eqref{eq:barpsi} and the \sPFC model follows in \ref{app:DAPFC}.

\subsection{Comparison of sPFC and sAPFC models with and without averaged density variations}
\label{app:DAPFC}
In the PFC model, the minimization of the free energy functional \eqref{eq:PFC_flat} can be achieved for either one single phase (disordered/liquid, stripe, triangular, \dots) or the coexistence of more than one phase \citep{Elder2002}. The latter is characterized by different arrangements described by different solutions of $\psi$, which may also have different $\overline\psi$. Such coexistence cannot be achieved in the APFC model where $\overline\psi$ is assumed to be constant. A model recovering the same coexistence condition has been proposed in \cite{Yeon2010}, where the average density $\overline{\psi}$ is allowed to vary spatially and evolved via $H^{-1}$ gradient flow (hereafter called D-APFC). While this approach better captures the phase transitions like melting, it introduces further approximations. In particular, it enforces that $\overline{\psi}$ is also slowly varying, similar to $\eta_n$ (we refer to \cite{Yeon2010} for further discussion). We found that, when looking at defects, the variation of $\overline{\psi}$ is significantly more pronounced in the APFC model with non-constant $\overline{\psi}$ than in the PFC models (in both the flat case and the surface formulations considered in this work), which is then better approximated by the APFC model with constant $\overline{\psi}$.

\begin{figure}
    \centering
    \includegraphics[width=0.85\linewidth]{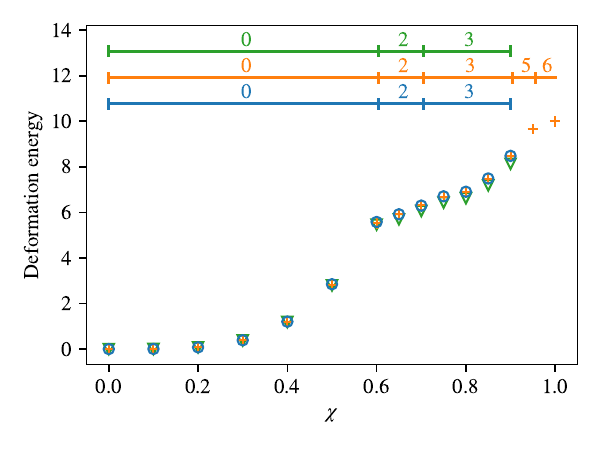}
    \caption{Deformation energy ($\calF_{\eta,\calS}-\calF_0$) and number of dislocations as a function of aspect ratio similarly to Fig.~\ref{fig:bump_pfc_apfc}. The results for the \sPFC (blue circles), \sAPFC (orange crosses), and D-APFC (green triangles) models are shown. Parameters: $A=4.0, B=0.03, C=-0.5,$ and $D=1/3$.}
    \label{fig:pfc_apfc_dapfc}
\end{figure}

To illustrate this in the context of the \sAPFC model presented in this work, we propose the study shown in Fig.~\ref{fig:bump_pfc_apfc} by comparing \sPFCnospace, \sAPFC and the surface formulation of the D-APFC model, the latter corresponding to solving the full system of Eqs. \eqref{eq:hapfc}--\eqref{eq:barpsi}. The results are illustrated in Fig.~\ref{fig:pfc_apfc_dapfc}. We see that for the same set of parameters entering the three models, a (slightly) better agreement is found between the \sPFC and \sAPFC models. However, with the \sAPFC model, melting for a larger aspect ratio is observed. We remark that the difference in the defect energy between the \sPFC model and the height formulation of the D-APFC model can be compensated by properly choosing different values of the parameter $A$ entering the two models. This can be exploited if targeting order-disorder / solid-liquid transition in the thin sheet, which is however not within the scope of the present investigation. It is worth mentioning that the evidence reported in this section also applies to the classical formulation of the models, \ie it does not depend on the surface parametrization chosen here.


\begin{thebibliography}{86}
\providecommand{\natexlab}[1]{#1}
\providecommand{\url}[1]{\texttt{#1}}
\expandafter\ifx\csname urlstyle\endcsname\relax
  \providecommand{\doi}[1]{\href{https://doi.org/#1}{doi: #1}}\else
  \providecommand{\doi}{doi: \begingroup \urlstyle{rm}\Url}\fi

\bibitem[Aland et~al.(2011)Aland, Lowengrub, and Voigt]{Aland2011}
Aland, S., Lowengrub, J., and Voigt, A.
\newblock {A continuum model of colloid-stabilized interfaces}.
\newblock \emph{Physics of Fluids}, 23\penalty0 (6):\penalty0 062103, 06 2011.
\newblock \href{http://dx.doi.org/10.1063/1.3584815}{\doi{10.1063/1.3584815}}.

\bibitem[Aland et~al.(2012{\natexlab{a}})Aland, Lowengrub, and
  Voigt]{Aland2012b}
Aland, S., Lowengrub, J., and Voigt, A.
\newblock Particles at fluid-fluid interfaces: A new
  navier-stokes-cahn-hilliard surface- phase-field-crystal model.
\newblock \emph{Phys. Rev. E}, 86:\penalty0 046321, Oct 2012{\natexlab{a}}.
\newblock
  \href{http://dx.doi.org/10.1103/PhysRevE.86.046321}{\doi{10.1103/PhysRevE.86.046321}}.

\bibitem[Aland et~al.(2012{\natexlab{b}})Aland, R\"{a}tz, R\"{o}ger, and
  Voigt]{Aland2012}
Aland, S., R\"{a}tz, A., R\"{o}ger, M., and Voigt, A.
\newblock Buckling instability of viral capsids—a continuum approach.
\newblock \emph{Multiscale Modeling \& Simulation}, 10\penalty0 (1):\penalty0
  82--110, 2012{\natexlab{b}}.
\newblock \href{http://dx.doi.org/10.1137/110834718}{\doi{10.1137/110834718}}.

\bibitem[Anderson et~al.(2017)Anderson, Hirth, and Lothe]{anderson2017}
Anderson, P., Hirth, J., and Lothe, J.
\newblock \emph{Theory of Dislocations}.
\newblock \href{http://www.worldcat.org/oclc/1132912878}{Cambridge University
  Press}, 2017.

\bibitem[Androulidakis et~al.(2018)Androulidakis, Zhang, Robertson, and
  Tawfick]{androulidakis2018tailoring}
Androulidakis, C., Zhang, K., Robertson, M., and Tawfick, S.
\newblock Tailoring the mechanical properties of 2d materials and
  heterostructures.
\newblock \emph{2D Materials}, 5\penalty0 (3):\penalty0 032005, 2018.
\newblock
  \href{http://dx.doi.org/10.1088/2053-1583/aac764}{\doi{10.1088/2053-1583/aac764}}.

\bibitem[Archer et~al.(2019)Archer, Ratliff, Rucklidge, and
  Subramanian]{Archer2019}
Archer, A.~J., Ratliff, D.~J., Rucklidge, A.~M., and Subramanian, P.
\newblock Deriving phase field crystal theory from dynamical density functional
  theory: Consequences of the approximations.
\newblock \emph{Physical Review E}, 100:\penalty0 022140, Aug 2019.
\newblock
  \href{http://dx.doi.org/10.1103/PhysRevE.100.022140}{\doi{10.1103/PhysRevE.100.022140}}.

\bibitem[Ariza and Ortiz(2010)]{Ariza2010}
Ariza, M.~P. and Ortiz, M.
\newblock Discrete dislocations in graphene.
\newblock \emph{Journal of the Mechanics and Physics of Solids}, 58\penalty0
  (5):\penalty0 710--734, 2010.
\newblock
  \href{http://dx.doi.org/10.1016/j.jmps.2010.02.008}{\doi{10.1016/j.jmps.2010.02.008}}.

\bibitem[Athreya et~al.(2006)Athreya, Goldenfeld, and Dantzig]{Athreya2006}
Athreya, B.~P., Goldenfeld, N., and Dantzig, J.~A.
\newblock Renormalization-group theory for the phase-field crystal equation.
\newblock \emph{Physical Review E}, 74\penalty0 (1):\penalty0 011601, 2006.
\newblock
  \href{http://dx.doi.org/10.1103/PhysRevE.74.011601}{\doi{10.1103/PhysRevE.74.011601}}.

\bibitem[Backofen et~al.(2010)Backofen, Voigt, and Witkowski]{Backofen2010}
Backofen, R., Voigt, A., and Witkowski, T.
\newblock Particles on curved surfaces: A dynamic approach by a
  phase-field-crystal model.
\newblock \emph{Physical Review E}, 81:\penalty0 025701, Feb 2010.
\newblock
  \href{http://dx.doi.org/10.1103/PhysRevE.81.025701}{\doi{10.1103/PhysRevE.81.025701}}.

\bibitem[Backofen et~al.(2011)Backofen, Gr{\"a}f, Potts, Praetorius, Voigt, and
  Witkowski]{Backofen2011}
Backofen, R., Gr{\"a}f, M., Potts, D., Praetorius, S., Voigt, A., and
  Witkowski, T.
\newblock A {{Continuous Approach}} to {{Discrete Ordering}} on $\mathbb{S}^2$.
\newblock \emph{Multiscale Modeling \& Simulation}, 9\penalty0 (1):\penalty0
  314--334, 2011.
\newblock \href{http://dx.doi.org/10.1137/100787532}{\doi{10.1137/100787532}}.

\bibitem[Backofen et~al.(2021)Backofen, Sahlmann, Willmann, and
  Voigt]{Backofen2021}
Backofen, R., Sahlmann, L., Willmann, A., and Voigt, A.
\newblock A comparison of different approaches to enforce lattice symmetry in
  two-dimensional crystals.
\newblock \emph{PAMM}, 20\penalty0 (1):\penalty0 e202000192, 2021.
\newblock
  \href{http://dx.doi.org/10.1002/pamm.202000192}{\doi{10.1002/pamm.202000192}}.

\bibitem[Bausch et~al.(2003)Bausch, Bowick, Cacciuto, Dinsmore, Hsu, Nelson,
  Nikolaides, Travesset, and Weitz]{Bausch2003}
Bausch, A.~R., Bowick, M.~J., Cacciuto, A., Dinsmore, A.~D., Hsu, M.~F.,
  Nelson, D.~R., Nikolaides, M.~G., Travesset, A., and Weitz, D.~A.
\newblock Grain boundary scars and spherical crystallography.
\newblock \emph{Science}, 299\penalty0 (5613):\penalty0 1716--1718, 2003.
\newblock
  \href{http://dx.doi.org/10.1126/science.1081160}{\doi{10.1126/science.1081160}}.

\bibitem[Benoit-Maréchal and Salvalaglio(2024)]{LBM_GE_2024gradient}
Benoit-Maréchal, L. and Salvalaglio, M.
\newblock Gradient elasticity in swift-hohenberg and phase-field crystal
  models, 2024.

\bibitem[Berry et~al.(2014)Berry, Provatas, Rottler, and Sinclair]{Berry2014}
Berry, J., Provatas, N., Rottler, J., and Sinclair, C.~W.
\newblock Phase field crystal modeling as a unified atomistic approach to
  defect dynamics.
\newblock \emph{Phys. Rev. B}, 89:\penalty0 214117, Jun 2014.
\newblock
  \href{http://dx.doi.org/10.1103/PhysRevB.89.214117}{\doi{10.1103/PhysRevB.89.214117}}.

\bibitem[Bulatov and Cai(2006)]{bulatov2006computer}
Bulatov, V. and Cai, W.
\newblock
  \emph{\href{https://global.oup.com/academic/product/computer-simulations-of-dislocations-9780199674060}{Computer
  simulations of dislocations}}, volume~3.
\newblock OUP Oxford, 2006.

\bibitem[Certaine(1960)]{Certaine1960}
Certaine, J.
\newblock The solution of ordinary differential equations with large time
  constants.
\newblock In \emph{Mathematical Methods for Digital Computers}, pages 128--132,
  New York, 1960. Wiley.

\bibitem[Cui et~al.(2020)Cui, Mukherjee, Sudeep, Colas, Najafi, Tam, Ajayan,
  Singh, Sun, and Filleter]{Cui2020}
Cui, T., Mukherjee, S., Sudeep, P.~M., Colas, G., Najafi, F., Tam, J., Ajayan,
  P.~M., Singh, C.~V., Sun, Y., and Filleter, T.
\newblock Fatigue of graphene.
\newblock \emph{Nature Materials}, 19\penalty0 (4):\penalty0 405--411, 2020.
\newblock
  \href{http://dx.doi.org/10.1038/s41563-019-0586-y}{\doi{10.1038/s41563-019-0586-y}}.

\bibitem[Dai et~al.(2016{\natexlab{a}})Dai, Xiang, and Srolovitz]{Dai2016}
Dai, S., Xiang, Y., and Srolovitz, D.~J.
\newblock Twisted {{Bilayer Graphene}}: {{Moir\'e}} with a {{Twist}}.
\newblock \emph{Nano Letters}, 16\penalty0 (9):\penalty0 5923--5927,
  2016{\natexlab{a}}.
\newblock
  \href{http://dx.doi.org/10.1021/acs.nanolett.6b02870}{\doi{10.1021/acs.nanolett.6b02870}}.

\bibitem[Dai et~al.(2016{\natexlab{b}})Dai, Xiang, and Srolovitz]{DaiPRB2016}
Dai, S., Xiang, Y., and Srolovitz, D.~J.
\newblock Structure and energetics of interlayer dislocations in bilayer
  graphene.
\newblock \emph{Phys. Rev. B}, 93:\penalty0 085410, Feb 2016{\natexlab{b}}.
\newblock
  \href{http://dx.doi.org/10.1103/PhysRevB.93.085410}{\doi{10.1103/PhysRevB.93.085410}}.

\bibitem[De~Donno et~al.(2023)De~Donno, Benoit-Mar\'echal, and
  Salvalaglio]{DeDonnoPRM2023}
De~Donno, M., Benoit-Mar\'echal, L., and Salvalaglio, M.
\newblock Amplitude expansion of the phase-field crystal model for complex
  crystal structures.
\newblock \emph{Physical Review Materials}, 7:\penalty0 033804, Mar 2023.
\newblock
  \href{http://dx.doi.org/10.1103/PhysRevMaterials.7.033804}{\doi{10.1103/PhysRevMaterials.7.033804}}.

\bibitem[De~Gennes and Prost(1993)]{de1993physics}
De~Gennes, P.-G. and Prost, J.
\newblock \emph{The physics of liquid crystals}.
\newblock Number~83.
  \href{https://global.oup.com/academic/product/the-physics-of-liquid-crystals-9780198517856?cc=de&lang=en&}{Oxford
  university press}, 1993.

\bibitem[Elder and Grant(2004)]{Elder2004}
Elder, K.~R. and Grant, M.
\newblock Modeling elastic and plastic deformations in nonequilibrium
  processing using phase field crystals.
\newblock \emph{Physical Review E}, 70\penalty0 (5):\penalty0 051605, 2004.
\newblock
  \href{http://dx.doi.org/10.1103/PhysRevE.70.051605}{\doi{10.1103/PhysRevE.70.051605}}.

\bibitem[Elder et~al.(2002)Elder, Katakowski, Haataja, and Grant]{Elder2002}
Elder, K.~R., Katakowski, M., Haataja, M., and Grant, M.
\newblock Modeling {{Elasticity}} in {{Crystal Growth}}.
\newblock \emph{Physical Review Letters}, 88\penalty0 (24):\penalty0 245701,
  2002.
\newblock
  \href{http://dx.doi.org/10.1103/PhysRevLett.88.245701}{\doi{10.1103/PhysRevLett.88.245701}}.

\bibitem[Elder et~al.(2007)Elder, Provatas, Berry, Stefanovic, and
  Grant]{Elder2007}
Elder, K.~R., Provatas, N., Berry, J., Stefanovic, P., and Grant, M.
\newblock Phase-field crystal modeling and classical density functional theory
  of freezing.
\newblock \emph{Physical Review B}, 75\penalty0 (6):\penalty0 064107, 2007.
\newblock
  \href{http://dx.doi.org/10.1103/PhysRevB.75.064107}{\doi{10.1103/PhysRevB.75.064107}}.

\bibitem[Elder et~al.(2010)Elder, Huang, and Provatas]{ElderPRE2010}
Elder, K.~R., Huang, Z.-F., and Provatas, N.
\newblock {Amplitude expansion of the binary phase-field-crystal model}.
\newblock \emph{Phys. Rev. E}, 81\penalty0 (1):\penalty0 011602, 2010.
\newblock
  \href{http://dx.doi.org/10.1103/PhysRevE.81.011602}{\doi{10.1103/PhysRevE.81.011602}}.

\bibitem[Elder et~al.(2021)Elder, Achim, Heinonen, Granato, Ying, and
  {Ala-Nissila}]{Elder2021}
Elder, K.~R., Achim, C.~V., Heinonen, V., Granato, E., Ying, S.~C., and
  {Ala-Nissila}, T.
\newblock Modeling buckling and topological defects in stacked two-dimensional
  layers of graphene and hexagonal boron nitride.
\newblock \emph{Physical Review Materials}, 5\penalty0 (3):\penalty0 034004,
  2021.
\newblock
  \href{http://dx.doi.org/10.1103/PhysRevMaterials.5.034004}{\doi{10.1103/PhysRevMaterials.5.034004}}.

\bibitem[Elder et~al.(2023)Elder, Huang, and Ala-Nissila]{Elder2023}
Elder, K.~R., Huang, Z.-F., and Ala-Nissila, T.
\newblock Moir\'e patterns and inversion boundaries in graphene/hexagonal boron
  nitride bilayers.
\newblock \emph{Physical Review Materials}, 7:\penalty0 024003, Feb 2023.
\newblock
  \href{http://dx.doi.org/10.1103/PhysRevMaterials.7.024003}{\doi{10.1103/PhysRevMaterials.7.024003}}.

\bibitem[Emmerich et~al.(2012)Emmerich, L{\"o}wen, Wittkowski, Gruhn, T{\'o}th,
  Tegze, and Gr{\'a}n{\'a}sy]{Emmerich2012}
Emmerich, H., L{\"o}wen, H., Wittkowski, R., Gruhn, T., T{\'o}th, G.~I., Tegze,
  G., and Gr{\'a}n{\'a}sy, L.
\newblock Phase-field-crystal models for condensed matter dynamics on atomic
  length and diffusive time scales: An overview.
\newblock \emph{Advances in Physics}, 61\penalty0 (6):\penalty0 665--743, 2012.
\newblock
  \href{http://dx.doi.org/10.1080/00018732.2012.737555}{\doi{10.1080/00018732.2012.737555}}.

\bibitem[Felton et~al.(2014)Felton, Tolley, Demaine, Rus, and Wood]{Felton2014}
Felton, S., Tolley, M., Demaine, E., Rus, D., and Wood, R.
\newblock A method for building self-folding machines.
\newblock \emph{Science}, 345\penalty0 (6197):\penalty0 644--646, 2014.
\newblock
  \href{http://dx.doi.org/10.1126/science.1252610}{\doi{10.1126/science.1252610}}.

\bibitem[Frigo and Johnson(2005)]{Frigo2005}
Frigo, M. and Johnson, S.
\newblock The {{Design}} and {{Implementation}} of {{FFTW3}}.
\newblock \emph{Proceedings of the IEEE}, 93\penalty0 (2):\penalty0 216--231,
  2005.
\newblock
  \href{http://dx.doi.org/10.1109/JPROC.2004.840301}{\doi{10.1109/JPROC.2004.840301}}.

\bibitem[Goldenfeld et~al.(2005)Goldenfeld, Athreya, and
  Dantzig]{Goldenfeld2005}
Goldenfeld, N., Athreya, B.~P., and Dantzig, J.~A.
\newblock Renormalization group approach to multiscale simulation of
  polycrystalline materials using the phase field crystal model.
\newblock \emph{Physical Review E}, 72\penalty0 (2):\penalty0 020601, 2005.
\newblock
  \href{http://dx.doi.org/10.1103/PhysRevE.72.020601}{\doi{10.1103/PhysRevE.72.020601}}.

\bibitem[Granato et~al.(2023)Granato, Elder, Ying, and
  Ala-Nissila]{Granato2023}
Granato, E., Elder, K.~R., Ying, S.~C., and Ala-Nissila, T.
\newblock Dynamics of fluctuations and thermal buckling in graphene from a
  phase-field crystal model.
\newblock \emph{Physical Review B}, 107:\penalty0 035428, Jan 2023.
\newblock
  \href{http://dx.doi.org/10.1103/PhysRevB.107.035428}{\doi{10.1103/PhysRevB.107.035428}}.

\bibitem[Guinea et~al.(2008)Guinea, Horovitz, and Le~Doussal]{Guinea2008}
Guinea, F., Horovitz, B., and Le~Doussal, P.
\newblock Gauge field induced by ripples in graphene.
\newblock \emph{Physical Review B}, 77\penalty0 (20):\penalty0 205421, 2008.
\newblock
  \href{http://dx.doi.org/10.1103/PhysRevB.77.205421}{\doi{10.1103/PhysRevB.77.205421}}.

\bibitem[Guo et~al.(2021)Guo, Hu, Liu, and Tian]{guo2021stacking}
Guo, H.-W., Hu, Z., Liu, Z.-B., and Tian, J.-G.
\newblock Stacking of 2d materials.
\newblock \emph{Advanced Functional Materials}, 31\penalty0 (4):\penalty0
  2007810, 2021.
\newblock
  \href{http://dx.doi.org/10.1002/adfm.202007810}{\doi{10.1002/adfm.202007810}}.

\bibitem[Heinonen et~al.(2014)Heinonen, Achim, Elder, Buyukdagli, and
  {Ala-Nissila}]{Heinonen2014}
Heinonen, V., Achim, C.~V., Elder, K.~R., Buyukdagli, S., and {Ala-Nissila}, T.
\newblock Phase-field-crystal models and mechanical equilibrium.
\newblock \emph{Physical Review E}, 89\penalty0 (3):\penalty0 032411, 2014.
\newblock
  \href{http://dx.doi.org/10.1103/PhysRevE.89.032411}{\doi{10.1103/PhysRevE.89.032411}}.

\bibitem[Heinonen et~al.(2016)Heinonen, Achim, Kosterlitz, Ying, Lowengrub, and
  {Ala-Nissila}]{Heinonen2016}
Heinonen, V., Achim, C.~V., Kosterlitz, J.~M., Ying, S.-C., Lowengrub, J., and
  {Ala-Nissila}, T.
\newblock Consistent {{Hydrodynamics}} for {{Phase Field Crystals}}.
\newblock \emph{Physical Review Letters}, 116\penalty0 (2):\penalty0 024303,
  2016.
\newblock
  \href{http://dx.doi.org/10.1103/PhysRevLett.116.024303}{\doi{10.1103/PhysRevLett.116.024303}}.

\bibitem[Helfrich(1973)]{Helfrich1973}
Helfrich, W.
\newblock Elastic properties of lipid bilayers: Theory and possible
  experiments.
\newblock \emph{Zeitschrift für Naturforschung C}, 28\penalty0
  (11-12):\penalty0 693--703, 1973.
\newblock
  \href{http://dx.doi.org/10.1515/znc-1973-11-1209}{\doi{10.1515/znc-1973-11-1209}}.

\bibitem[Hexemer et~al.(2007)Hexemer, Vitelli, Kramer, and
  Fredrickson]{Hexemer2007}
Hexemer, A., Vitelli, V., Kramer, E.~J., and Fredrickson, G.~H.
\newblock Monte {{Carlo}} study of crystalline order and defects on weakly
  curved surfaces.
\newblock \emph{Physical Review E}, 76\penalty0 (5):\penalty0 051604, 2007.
\newblock
  \href{http://dx.doi.org/10.1103/PhysRevE.76.051604}{\doi{10.1103/PhysRevE.76.051604}}.

\bibitem[Hirvonen et~al.(2017)Hirvonen, Fan, Ervasti, Harju, Elder, and
  {Ala-Nissila}]{Hirvonen2017}
Hirvonen, P., Fan, Z., Ervasti, M.~M., Harju, A., Elder, K.~R., and
  {Ala-Nissila}, T.
\newblock Energetics and structure of grain boundary triple junctions in
  graphene.
\newblock \emph{Scientific Reports}, 7\penalty0 (1):\penalty0 4754, 2017.
\newblock
  \href{http://dx.doi.org/10.1038/s41598-017-04852-w}{\doi{10.1038/s41598-017-04852-w}}.

\bibitem[Hu et~al.(2012)Hu, Meng, Li, and Ibekwe]{Hu2012review}
Hu, J., Meng, H., Li, G., and Ibekwe, S.~I.
\newblock A review of stimuli-responsive polymers for smart textile
  applications.
\newblock \emph{Smart Materials and Structures}, 21\penalty0 (5):\penalty0
  053001, 2012.
\newblock
  \href{http://dx.doi.org/10.1088/0964-1726/21/5/053001}{\doi{10.1088/0964-1726/21/5/053001}}.

\bibitem[H{\"u}ter et~al.(2016)H{\"u}ter, Fri{\'a}k, Weikamp, Neugebauer,
  Goldenfeld, Svendsen, and Spatschek]{Huter2016}
H{\"u}ter, C., Fri{\'a}k, M., Weikamp, M., Neugebauer, J., Goldenfeld, N.,
  Svendsen, B., and Spatschek, R.
\newblock Nonlinear elastic effects in phase field crystal and amplitude
  equations: {{Comparison}} to ab initio simulations of bcc metals and
  graphene.
\newblock \emph{Physical Review B}, 93\penalty0 (21):\penalty0 214105, 2016.
\newblock
  \href{http://dx.doi.org/10.1103/PhysRevB.93.214105}{\doi{10.1103/PhysRevB.93.214105}}.

\bibitem[Jeong et~al.(2008)Jeong, Ihm, and Lee]{Jeong2008}
Jeong, B.~W., Ihm, J., and Lee, G.-D.
\newblock Stability of dislocation defect with two pentagon-heptagon pairs in
  graphene.
\newblock \emph{Physical Review B}, 78\penalty0 (16):\penalty0 165403, 2008.
\newblock
  \href{http://dx.doi.org/10.1103/PhysRevB.78.165403}{\doi{10.1103/PhysRevB.78.165403}}.

\bibitem[Jreidini et~al.(2021)Jreidini, Pinomaa, Wiezorek, McKeown, Laukkanen,
  and Provatas]{Jreidini2021}
Jreidini, P., Pinomaa, T., Wiezorek, J. M.~K., McKeown, J.~T., Laukkanen, A.,
  and Provatas, N.
\newblock Orientation gradients in rapidly solidified pure aluminum thin films:
  Comparison of experiments and phase-field crystal simulations.
\newblock \emph{Phys. Rev. Lett.}, 127:\penalty0 205701, Nov 2021.
\newblock
  \href{http://dx.doi.org/10.1103/PhysRevLett.127.205701}{\doi{10.1103/PhysRevLett.127.205701}}.

\bibitem[Khanra et~al.(2022)Khanra, Jia, Mitchell, Balchunas, Pelcovits,
  Powers, Dogic, and Sharma]{khanra2022controlling}
Khanra, A., Jia, L.~L., Mitchell, N.~P., Balchunas, A., Pelcovits, R.~A.,
  Powers, T.~R., Dogic, Z., and Sharma, P.
\newblock Controlling the shape and topology of two-component colloidal
  membranes.
\newblock \emph{Proceedings of the National Academy of Sciences}, 119\penalty0
  (32):\penalty0 e2204453119, 2022.
\newblock
  \href{http://dx.doi.org/10.1073/pnas.2204453119}{\doi{10.1073/pnas.2204453119}}.

\bibitem[K{\"o}hler et~al.(2016)K{\"o}hler, Backofen, and Voigt]{Kohler2016}
K{\"o}hler, C., Backofen, R., and Voigt, A.
\newblock Stress {{Induced Branching}} of {{Growing Crystals}} on {{Curved
  Surfaces}}.
\newblock \emph{Physical Review Letters}, 116\penalty0 (13):\penalty0 135502,
  2016.
\newblock
  \href{http://dx.doi.org/10.1103/PhysRevLett.116.135502}{\doi{10.1103/PhysRevLett.116.135502}}.

\bibitem[Kubin(2013)]{kubin2013dislocations}
Kubin, L.
\newblock
  \emph{\href{https://global.oup.com/academic/product/dislocations-mesoscale-simulations-and-plastic-flow-9780198525011?cc=de&lang=en&}{Dislocations,
  mesoscale simulations and plastic flow}}, volume~5.
\newblock Oxford University Press, USA, 2013.

\bibitem[Lee et~al.(2014)Lee, Yoon, He, Robertson, and Warner]{C4NR04718D}
Lee, G.-D., Yoon, E., He, K., Robertson, A.~W., and Warner, J.~H.
\newblock Detailed formation processes of stable dislocations in graphene.
\newblock \emph{Nanoscale}, 6:\penalty0 14836--14844, 2014.
\newblock
  \href{http://dx.doi.org/10.1039/C4NR04718D}{\doi{10.1039/C4NR04718D}}.

\bibitem[Lehtinen et~al.(2013)Lehtinen, Kurasch, Krasheninnikov, and
  Kaiser]{Lehtinen2013}
Lehtinen, O., Kurasch, S., Krasheninnikov, A.~V., and Kaiser, U.
\newblock Atomic scale study of the life cycle of a dislocation in graphene
  from birth to annihilation.
\newblock \emph{Nature Communications}, 4:\penalty0 2098, 2013.
\newblock
  \href{http://dx.doi.org/10.1038/ncomms3098}{\doi{10.1038/ncomms3098}}.

\bibitem[Liu et~al.(2011)Liu, Gajewski, Pao, and Chang]{Liu2011}
Liu, T.-H., Gajewski, G., Pao, C.-W., and Chang, C.-C.
\newblock Structure, energy, and structural transformations of graphene grain
  boundaries from atomistic simulations.
\newblock \emph{Carbon}, 49\penalty0 (7):\penalty0 2306--2317, 2011.
\newblock
  \href{http://dx.doi.org/10.1016/j.carbon.2011.01.063}{\doi{10.1016/j.carbon.2011.01.063}}.

\bibitem[Michael~te Vrugt and Wittkowski(2020)]{teVrugt2020}
Michael~te Vrugt, H.~L. and Wittkowski, R.
\newblock Classical dynamical density functional theory: from fundamentals to
  applications.
\newblock \emph{Advances in Physics}, 69\penalty0 (2):\penalty0 121--247, 2020.
\newblock
  \href{http://dx.doi.org/10.1080/00018732.2020.1854965}{\doi{10.1080/00018732.2020.1854965}}.

\bibitem[Mkhonta et~al.(2013)Mkhonta, Elder, and Huang]{Mkhonta2013}
Mkhonta, S.~K., Elder, K.~R., and Huang, Z.-F.
\newblock {Exploring the Complex World of Two-Dimensional Ordering with Three
  Modes}.
\newblock \emph{Physical Review Letters}, 111:\penalty0 035501, 2013.
\newblock
  \href{http://dx.doi.org/10.1103/PhysRevLett.111.035501}{\doi{10.1103/PhysRevLett.111.035501}}.

\bibitem[Molaei et~al.(2021)Molaei, Younas, and Rezakazemi]{Molaei2021}
Molaei, M.~J., Younas, M., and Rezakazemi, M.
\newblock A comprehensive review on recent advances in two-dimensional (2d)
  hexagonal boron nitride.
\newblock \emph{ACS Applied Electronic Materials}, 3\penalty0 (12):\penalty0
  5165--5187, 12 2021.
\newblock
  \href{http://dx.doi.org/10.1021/acsaelm.1c00720}{\doi{10.1021/acsaelm.1c00720}}.

\bibitem[Nitschke et~al.(2020)Nitschke, Reuther, and Voigt]{Nitschke2020}
Nitschke, I., Reuther, S., and Voigt, A.
\newblock Liquid crystals on deformable surfaces.
\newblock \emph{Proceedings of the Royal Society A: Mathematical, Physical and
  Engineering Sciences}, 476\penalty0 (2241):\penalty0 20200313, 2020.
\newblock
  \href{http://dx.doi.org/10.1098/rspa.2020.0313}{\doi{10.1098/rspa.2020.0313}}.

\bibitem[Nitschke et~al.(2024)Nitschke, Sadik, and Voigt]{Nitschke2024}
Nitschke, I., Sadik, S., and Voigt, A.
\newblock {Tangential tensor fields on deformable surfaces--how to derive
  consistent L2-gradient flows}.
\newblock \emph{IMA Journal of Applied Mathematics}, page hxae006, 02 2024.
\newblock
  \href{http://dx.doi.org/10.1093/imamat/hxae006}{\doi{10.1093/imamat/hxae006}}.

\bibitem[Pereira et~al.(2010)Pereira, Castro~Neto, Liang, and
  Mahadevan]{Pereira2010}
Pereira, V.~M., Castro~Neto, A.~H., Liang, H.~Y., and Mahadevan, L.
\newblock Geometry, {{Mechanics}}, and {{Electronics}} of {{Singular
  Structures}} and {{Wrinkles}} in {{Graphene}}.
\newblock \emph{Physical Review Letters}, 105\penalty0 (15):\penalty0 156603,
  2010.
\newblock
  \href{http://dx.doi.org/10.1103/PhysRevLett.105.156603}{\doi{10.1103/PhysRevLett.105.156603}}.

\bibitem[Pope(1963)]{Pope1963}
Pope, D.~A.
\newblock An exponential method of numerical integration of ordinary
  differential equations.
\newblock \emph{Commun. ACM}, 6\penalty0 (8):\penalty0 491–493, aug 1963.
\newblock
  \href{http://dx.doi.org/10.1145/366707.367592}{\doi{10.1145/366707.367592}}.

\bibitem[Praetorius et~al.(2019)Praetorius, Salvalaglio, and
  Voigt]{Praetorius2019}
Praetorius, S., Salvalaglio, M., and Voigt, A.
\newblock {An efficient numerical framework for the amplitude expansion of the
  phase-field crystal model}.
\newblock \emph{Model. Simul. Mater. Sci. Eng.}, 27\penalty0 (4):\penalty0
  044004, 2019.
\newblock
  \href{http://dx.doi.org/10.1088/1361-651X/ab1508}{\doi{10.1088/1361-651X/ab1508}}.

\bibitem[Rollett et~al.(2015)Rollett, Rohrer, and
  Suter]{rollett2015understanding}
Rollett, A., Rohrer, G., and Suter, R.
\newblock Understanding materials microstructure and behavior at the mesoscale.
\newblock \emph{MRS Bulletin}, 40\penalty0 (11):\penalty0 951--960, 2015.
\newblock
  \href{http://dx.doi.org/10.1557/mrs.2015.262}{\doi{10.1557/mrs.2015.262}}.

\bibitem[Roychowdhury and Gupta(2018)]{Roy2018}
Roychowdhury, A. and Gupta, A.
\newblock On {{Structured Surfaces}} with {{Defects}}: {{Geometry}}, {{Strain
  Incompatibility}}, {{Stress Field}}, and {{Natural Shapes}}.
\newblock \emph{Journal of Elasticity}, 131\penalty0 (2):\penalty0 239--276,
  2018.
\newblock
  \href{http://dx.doi.org/10.1007/s10659-017-9654-1}{\doi{10.1007/s10659-017-9654-1}}.

\bibitem[Salvalaglio and Elder(2022)]{Salvalaglio2022ov}
Salvalaglio, M. and Elder, K.~R.
\newblock Coarse-grained modeling of crystals by the amplitude expansion of the
  phase-field crystal model: An overview.
\newblock \emph{Modelling and Simulation in Materials Science and Engineering},
  30\penalty0 (5):\penalty0 053001, 2022.
\newblock
  \href{http://dx.doi.org/10.1088/1361-651X/ac681e}{\doi{10.1088/1361-651X/ac681e}}.

\bibitem[Salvalaglio et~al.(2019)Salvalaglio, Voigt, and
  Elder]{Salvalaglio2019}
Salvalaglio, M., Voigt, A., and Elder, K.~R.
\newblock Closing the gap between atomic-scale lattice deformations and
  continuum elasticity.
\newblock \emph{npj Computational Materials}, 5\penalty0 (1):\penalty0 1--9,
  2019.
\newblock
  \href{http://dx.doi.org/10.1038/s41524-019-0185-0}{\doi{10.1038/s41524-019-0185-0}}.

\bibitem[Salvalaglio et~al.(2020)Salvalaglio, Angheluta, Huang, Voigt, Elder,
  and Vi{\~n}als]{Salvalaglio2020plastic}
Salvalaglio, M., Angheluta, L., Huang, Z.-F., Voigt, A., Elder, K.~R., and
  Vi{\~n}als, J.
\newblock A coarse-grained phase-field crystal model of plastic motion.
\newblock \emph{Journal of the Mechanics and Physics of Solids}, 137:\penalty0
  103856, 2020.
\newblock
  \href{http://dx.doi.org/10.1016/j.jmps.2019.103856}{\doi{10.1016/j.jmps.2019.103856}}.

\bibitem[Salvalaglio et~al.(2021)Salvalaglio, Voigt, Huang, and
  Elder]{Salvalaglio2021}
Salvalaglio, M., Voigt, A., Huang, Z.-F., and Elder, K.~R.
\newblock Mesoscale {{Defect Motion}} in {{Binary Systems}}: {{Effects}} of
  {{Compositional Strain}} and {{Cottrell Atmospheres}}.
\newblock \emph{Physical Review Letters}, 126\penalty0 (18):\penalty0 185502,
  2021.
\newblock
  \href{http://dx.doi.org/10.1103/PhysRevLett.126.185502}{\doi{10.1103/PhysRevLett.126.185502}}.

\bibitem[Seung and Nelson(1988)]{Seung1988}
Seung, H.~S. and Nelson, D.~R.
\newblock Defects in flexible membranes with crystalline order.
\newblock \emph{Physical Review A}, 38\penalty0 (2):\penalty0 1005--1018, 1988.
\newblock
  \href{http://dx.doi.org/10.1103/PhysRevA.38.1005}{\doi{10.1103/PhysRevA.38.1005}}.

\bibitem[Singh et~al.(2022{\natexlab{a}})Singh, Pandey, and
  Gupta]{singh2022interaction}
Singh, M., Pandey, A., and Gupta, A.
\newblock Interaction of a defect with the reference curvature of an elastic
  surface.
\newblock \emph{Soft Matter}, 18\penalty0 (15):\penalty0 2979--2991,
  2022{\natexlab{a}}.
\newblock
  \href{http://dx.doi.org/10.1039/D2SM00126H}{\doi{10.1039/D2SM00126H}}.

\bibitem[Singh et~al.(2022{\natexlab{b}})Singh, Roychowdhury, and
  Gupta]{Singh2022}
Singh, M., Roychowdhury, A., and Gupta, A.
\newblock Defects and metric anomalies in föppl–von kármán surfaces.
\newblock \emph{Proceedings of the Royal Society A: Mathematical, Physical and
  Engineering Sciences}, 478\penalty0 (2262):\penalty0 20210829,
  2022{\natexlab{b}}.
\newblock
  \href{http://dx.doi.org/10.1098/rspa.2021.0829}{\doi{10.1098/rspa.2021.0829}}.

\bibitem[Skaugen et~al.(2018{\natexlab{a}})Skaugen, Angheluta, and
  Vi{\~n}als]{Skaugen2018}
Skaugen, A., Angheluta, L., and Vi{\~n}als, J.
\newblock Separation of {{Elastic}} and {{Plastic Timescales}} in a {{Phase
  Field Crystal Model}}.
\newblock \emph{Physical Review Letters}, 121\penalty0 (25):\penalty0 255501,
  2018{\natexlab{a}}.
\newblock
  \href{http://dx.doi.org/10.1103/PhysRevLett.121.255501}{\doi{10.1103/PhysRevLett.121.255501}}.

\bibitem[Skaugen et~al.(2018{\natexlab{b}})Skaugen, Angheluta, and
  Vi{\~n}als]{Skaugen2018a}
Skaugen, A., Angheluta, L., and Vi{\~n}als, J.
\newblock Dislocation dynamics and crystal plasticity in the phase-field
  crystal model.
\newblock \emph{Physical Review B}, 97\penalty0 (5):\penalty0 054113,
  2018{\natexlab{b}}.
\newblock
  \href{http://dx.doi.org/10.1103/PhysRevB.97.054113}{\doi{10.1103/PhysRevB.97.054113}}.

\bibitem[Skogvoll et~al.(2021)Skogvoll, Skaugen, and Angheluta]{Skogvoll2021}
Skogvoll, V., Skaugen, A., and Angheluta, L.
\newblock Stress in ordered systems: {{Ginzburg-Landau-type}} density field
  theory.
\newblock \emph{Physical Review B}, 103\penalty0 (22):\penalty0 224107, 2021.
\newblock
  \href{http://dx.doi.org/10.1103/PhysRevB.103.224107}{\doi{10.1103/PhysRevB.103.224107}}.

\bibitem[Skogvoll et~al.(2022{\natexlab{a}})Skogvoll, Angheluta, Skaugen,
  Salvalaglio, and Vi{\~n}als]{Skogvoll2022pfc}
Skogvoll, V., Angheluta, L., Skaugen, A., Salvalaglio, M., and Vi{\~n}als, J.
\newblock A phase field crystal theory of the kinematics of dislocation lines.
\newblock \emph{Journal of the Mechanics and Physics of Solids}, 166:\penalty0
  104932, 2022{\natexlab{a}}.
\newblock
  \href{http://dx.doi.org/10.1016/j.jmps.2022.104932}{\doi{10.1016/j.jmps.2022.104932}}.

\bibitem[Skogvoll et~al.(2022{\natexlab{b}})Skogvoll, Salvalaglio, and
  Angheluta]{Skogvoll2022}
Skogvoll, V., Salvalaglio, M., and Angheluta, L.
\newblock Hydrodynamic phase field crystal approach to interfaces,
  dislocations, and multi-grain networks.
\newblock \emph{Modelling and Simulation in Materials Science and Engineering},
  30\penalty0 (8):\penalty0 084002, 2022{\natexlab{b}}.
\newblock
  \href{http://dx.doi.org/10.1088/1361-651X/ac9493}{\doi{10.1088/1361-651X/ac9493}}.

\bibitem[Skogvoll et~al.(2023)Skogvoll, R{\o}nning, Salvalaglio, and
  Angheluta]{Skogvoll2023}
Skogvoll, V., R{\o}nning, J., Salvalaglio, M., and Angheluta, L.
\newblock A unified field theory of topological defects and non-linear local
  excitations.
\newblock \emph{npj Computational Materials}, 9\penalty0 (1):\penalty0 122,
  2023.
\newblock
  \href{http://dx.doi.org/10.1038/s41524-023-01077-6}{\doi{10.1038/s41524-023-01077-6}}.

\bibitem[Stefanovic et~al.(2006)Stefanovic, Haataja, and
  Provatas]{Stefanovic2006}
Stefanovic, P., Haataja, M., and Provatas, N.
\newblock Phase-{{Field Crystals}} with {{Elastic Interactions}}.
\newblock \emph{Physical Review Letters}, 96\penalty0 (22):\penalty0 225504,
  2006.
\newblock
  \href{http://dx.doi.org/10.1103/PhysRevLett.96.225504}{\doi{10.1103/PhysRevLett.96.225504}}.

\bibitem[Sydney~Gladman et~al.(2016)Sydney~Gladman, Matsumoto, Nuzzo,
  Mahadevan, and Lewis]{SydneyGladman2016}
Sydney~Gladman, A., Matsumoto, E.~A., Nuzzo, R.~G., Mahadevan, L., and Lewis,
  J.~A.
\newblock Biomimetic {{4D}} printing.
\newblock \emph{Nature Materials}, 15\penalty0 (4):\penalty0 413--418, 2016.
\newblock \href{http://dx.doi.org/10.1038/nmat4544}{\doi{10.1038/nmat4544}}.

\bibitem[te~Vrugt and Wittkowski(2022)]{teVrugt2023}
te~Vrugt, M. and Wittkowski, R.
\newblock Perspective: New directions in dynamical density functional theory.
\newblock \emph{Journal of Physics: Condensed Matter}, 35\penalty0
  (4):\penalty0 041501, dec 2022.
\newblock
  \href{http://dx.doi.org/10.1088/1361-648X/ac8633}{\doi{10.1088/1361-648X/ac8633}}.
\newblock URL \url{https://dx.doi.org/10.1088/1361-648X/ac8633}.

\bibitem[{Torkaman-Asadi} and Kouchakzadeh(2022)]{Torkaman-Asadi2022}
{Torkaman-Asadi}, M.~A. and Kouchakzadeh, M.~A.
\newblock Atomistic simulations of mechanical properties and fracture of
  graphene: {{A}} review.
\newblock \emph{Computational Materials Science}, 210:\penalty0 111457, 2022.
\newblock
  \href{http://dx.doi.org/10.1016/j.commatsci.2022.111457}{\doi{10.1016/j.commatsci.2022.111457}}.

\bibitem[T{\'o}th et~al.(2013)T{\'o}th, Gr{\'a}n{\'a}sy, and Tegze]{Toth2013}
T{\'o}th, G.~I., Gr{\'a}n{\'a}sy, L., and Tegze, G.
\newblock Nonlinear hydrodynamic theory of crystallization.
\newblock \emph{Journal of Physics: Condensed Matter}, 26\penalty0
  (5):\penalty0 055001, 2013.
\newblock
  \href{http://dx.doi.org/10.1088/0953-8984/26/5/055001}{\doi{10.1088/0953-8984/26/5/055001}}.

\bibitem[van Teeffelen et~al.(2009)van Teeffelen, Backofen, Voigt, and
  L{\"o}wen]{Teeffelen2009}
van Teeffelen, S., Backofen, R., Voigt, A., and L{\"o}wen, H.
\newblock Derivation of the phase-field-crystal model for colloidal
  solidification.
\newblock \emph{Physical Review E}, 79\penalty0 (5):\penalty0 051404, 2009.
\newblock
  \href{http://dx.doi.org/10.1103/PhysRevE.79.051404}{\doi{10.1103/PhysRevE.79.051404}}.

\bibitem[{Vitelli}(2006)]{Vitelli2006a}
{Vitelli}, V.
\newblock
  \emph{{\href{https://www.yumpu.com/en/document/read/8163550/crystals-liquid-crystals-and-superfluid-helium-on-curved-surfaces}{Crystals,
  liquid crystals and superfluid helium on curved surfaces}}}.
\newblock PhD thesis, Harvard University, Massachusetts, Jan. 2006.

\bibitem[Wang et~al.(2018)Wang, Liu, and Huang]{Wang2018}
Wang, Z.~L., Liu, Z., and Huang, Z.~F.
\newblock {Angle-adjustable density field formulation for the modeling of
  crystalline microstructure}.
\newblock \emph{Physical Review B}, 97:\penalty0 180102, 2018.
\newblock
  \href{http://dx.doi.org/10.1103/PhysRevB.97.180102}{\doi{10.1103/PhysRevB.97.180102}}.

\bibitem[Warner et~al.(2012)Warner, Margine, Mukai, Robertson, Giustino, and
  Kirkland]{Warner2012}
Warner, J.~H., Margine, E.~R., Mukai, M., Robertson, A.~W., Giustino, F., and
  Kirkland, A.~I.
\newblock Dislocation-{{Driven Deformations}} in {{Graphene}}.
\newblock \emph{Science}, 337\penalty0 (6091):\penalty0 209--212, 2012.
\newblock
  \href{http://dx.doi.org/10.1126/science.1217529}{\doi{10.1126/science.1217529}}.

\bibitem[Witten(2007)]{Witten2007}
Witten, T.~A.
\newblock Stress focusing in elastic sheets.
\newblock \emph{Reviews of Modern Physics}, 79\penalty0 (2):\penalty0 643--675,
  2007.
\newblock
  \href{http://dx.doi.org/10.1103/RevModPhys.79.643}{\doi{10.1103/RevModPhys.79.643}}.

\bibitem[Wu et~al.(2010)Wu, Plapp, and Voorhees]{Wu2010}
Wu, K.-A., Plapp, M., and Voorhees, P.~W.
\newblock {Controlling crystal symmetries in phase-field crystal models.}
\newblock \emph{Journal of Physics: Condensed Matter}, 22:\penalty0 364102,
  2010.
\newblock
  \href{http://dx.doi.org/10.1088/0953-8984/22/36/364102}{\doi{10.1088/0953-8984/22/36/364102}}.

\bibitem[Yeon et~al.(2010)Yeon, Huang, Elder, and Thornton]{Yeon2010}
Yeon, D.-H., Huang, Z.-F., Elder, K., and Thornton, K.
\newblock Density-amplitude formulation of the phase-field crystal model for
  two-phase coexistence in two and three dimensions.
\newblock \emph{Philosophical Magazine}, 90\penalty0 (1-4):\penalty0 237--263,
  2010.
\newblock
  \href{http://dx.doi.org/10.1080/14786430903164572}{\doi{10.1080/14786430903164572}}.

\bibitem[Zhang et~al.(2014{\natexlab{a}})Zhang, Li, and Gao]{Zhang2014a}
Zhang, T., Li, X., and Gao, H.
\newblock Defects controlled wrinkling and topological design in graphene.
\newblock \emph{Journal of the Mechanics and Physics of Solids}, 67:\penalty0
  2--13, 2014{\natexlab{a}}.
\newblock
  \href{http://dx.doi.org/10.1016/j.jmps.2014.02.005}{\doi{10.1016/j.jmps.2014.02.005}}.

\bibitem[Zhang et~al.(2014{\natexlab{b}})Zhang, Li, and Gao]{Zhang2014b}
Zhang, T., Li, X., and Gao, H.
\newblock Designing graphene structures with controlled distributions of
  topological defects: A case study of toughness enhancement in graphene ruga.
\newblock \emph{Extreme Mechanics Letters}, 1:\penalty0 3--8,
  2014{\natexlab{b}}.
\newblock
  \href{http://dx.doi.org/https://doi.org/10.1016/j.eml.2014.12.007}{\doi{https://doi.org/10.1016/j.eml.2014.12.007}}.

\end{thebibliography}
\end{document}